\documentclass[preprint,3p,times,12pt]{elsarticle}

\usepackage{amsmath, amssymb, pifont} 
\usepackage{graphicx}         
\usepackage{lineno}           
\usepackage{hyperref}         
\usepackage{xcolor}           
\usepackage{booktabs}         
\usepackage{multirow}         
\usepackage{float}            
\usepackage{tabularx}         
\usepackage{tikz}             
\usepackage{rotating}         
\usepackage[T1]{fontenc}      
\usepackage{subcaption}
\usepackage{natbib}

\hypersetup{
	colorlinks   = true,
	urlcolor     = blue,
	linkcolor    = blue,
	citecolor    = red
}
\newcolumntype{Y}{>{\centering\arraybackslash}X}
\AtBeginDocument{\DeclareMathAlphabet{\mathbf}{OT1}{cmr}{bx}{n}}

\usepackage{xcolor}
\definecolor{newcolor}{rgb}{.8,.349,.1}

\usepackage{tikz}
\usepackage{xcolor}

\definecolor{skyblue}{rgb}{0.26666666666666666, 0.4470588235294118, 0.7686274509803922}
\definecolor{lightgreen}{rgb}{0.4392156862745098, 0.6784313725490196, 0.2784313725490196}

\usetikzlibrary{shapes.geometric}

\newcommand{\greencircle}{\protect\tikz[baseline=-0.6ex]\protect\node[fill=lightgreen, circle, inner sep=1.5pt] {};}
\newcommand{\redcircle}{\protect\tikz[baseline=-0.6ex]\protect\node[fill=red, circle, inner sep=1.5pt] {};}

\definecolor{darkgreen}{rgb}{0.0, 0.5, 0.0}
\definecolor{darkred}{rgb}{0.55, 0.0, 0.0}

\usepackage{gensymb}
\usepackage{makecell}
\usepackage[linesnumbered,ruled,vlined]{algorithm2e}

\journal{Elsevier} 

\begin{document}

\begin{frontmatter}



\title{A fully differentiable GNN-based PDE Solver: With Applications to Poisson and Navier-Stokes Equations}

\author[inst1,inst2]{Tianyu Li}
\author[inst1,inst2]{Yiye Zou}
\author[inst3]{Shufan Zou}
\author[inst1,inst2]{Xinghua Chang \corref{mycorrespondingauthor}}
\cortext[mycorrespondingauthor]{Corresponding author}
\ead{cxh_cardc@126.com}
\author[inst1]{Laiping Zhang}
\author[inst4]{Xiaogang Deng}

\affiliation[inst1]{organization={National key laboratory of fundamental algorithms and models for engineering simulation},
	addressline={Sichuan University},
	city={Chengdu},
	postcode={610207},
	country={China}}
\affiliation[inst2]{organization={School of Computer Science},
	addressline={Sichuan University},
	city={Chengdu},
	postcode={610065},
	country={China}}
\affiliation[inst3]{organization={College of Aerospace Science and Engineering},
	addressline={National University of Defense Technology},
	city={Changsha},
	postcode={410000},
	country={China}}
\affiliation[inst4]{organization={Academy of Military Sciences},
	city={Beijing},
	postcode={100190},
	country={China}}

\begin{abstract}
In this study, we present a novel computational framework that integrates the finite volume method with graph neural networks to address the challenges in Physics-Informed Neural Networks(PINNs). Our approach leverages the flexibility of graph neural networks to adapt to various types of two-dimensional unstructured grids, enhancing the model's applicability across different physical equations and boundary conditions. The core innovation lies in the development of an unsupervised training algorithm that utilizes GPU parallel computing to implement a fully differentiable finite volume method discretization process. This method includes differentiable integral and gradient reconstruction algorithms, enabling the model to directly solve partial-differential equations(PDEs) during training without the need for pre-computed data. Our results demonstrate the model's superior mesh generalization and its capability to handle multiple boundary conditions simultaneously, significantly boosting its generalization capabilities. The proposed method not only shows potential for extensive applications in CFD but also establishes a new paradigm for integrating traditional numerical methods with deep learning technologies, offering a robust platform for solving complex physical problems.
\end{abstract}




\end{frontmatter}


\section{Introduction}\label{sec:Introduction}

Fluid dynamics is concerned with the motion and interactions of fluids in various physical scenarios, such as the behavior of air or water. To describe and simulate fluid problems, it is often necessary to solve Partial Differential Equations (PDEs). However, traditional numerical methods such as Finite Difference Method, Finite Element Method, and Finite Volume Method face significant computational efficiency challenges when dealing with complex geometries, multi-scale and multi-physics phenomena, and boundary conditions. This is primarily due to the substantial computational overhead caused by high-quality grid generation and the coupling nonlinear effects of various physical quantities, and their simulation accuracy also tends to decrease as the complexity of the physical scenarios increases. In recent years, deep neural network-based methods, such as the Physics-Informed Neural Network (PINNs)\cite{raissi_physics_2017,rao_physics-informed_2020,jin_nsfnets_2021} have achieved significant results in solving PDEs, including the Navier-Stokes equations. These methods rely on automatic differentiation (AD) in the deep learning framework to propagate physical laws within the neural network. Some work also involves applying traditional numerical methods for differential approximations, such as combining finite differences \cite{wandel2020learning,wandel_teaching_2021,gao_phygeonet_2021} and finite element methods\cite{gao_physics-informed_2022,xu2021conditionally}. 

\begin{figure}[!htbp]
\centering
\begin{minipage}{\linewidth}\centering
    \includegraphics[width=1\textwidth]{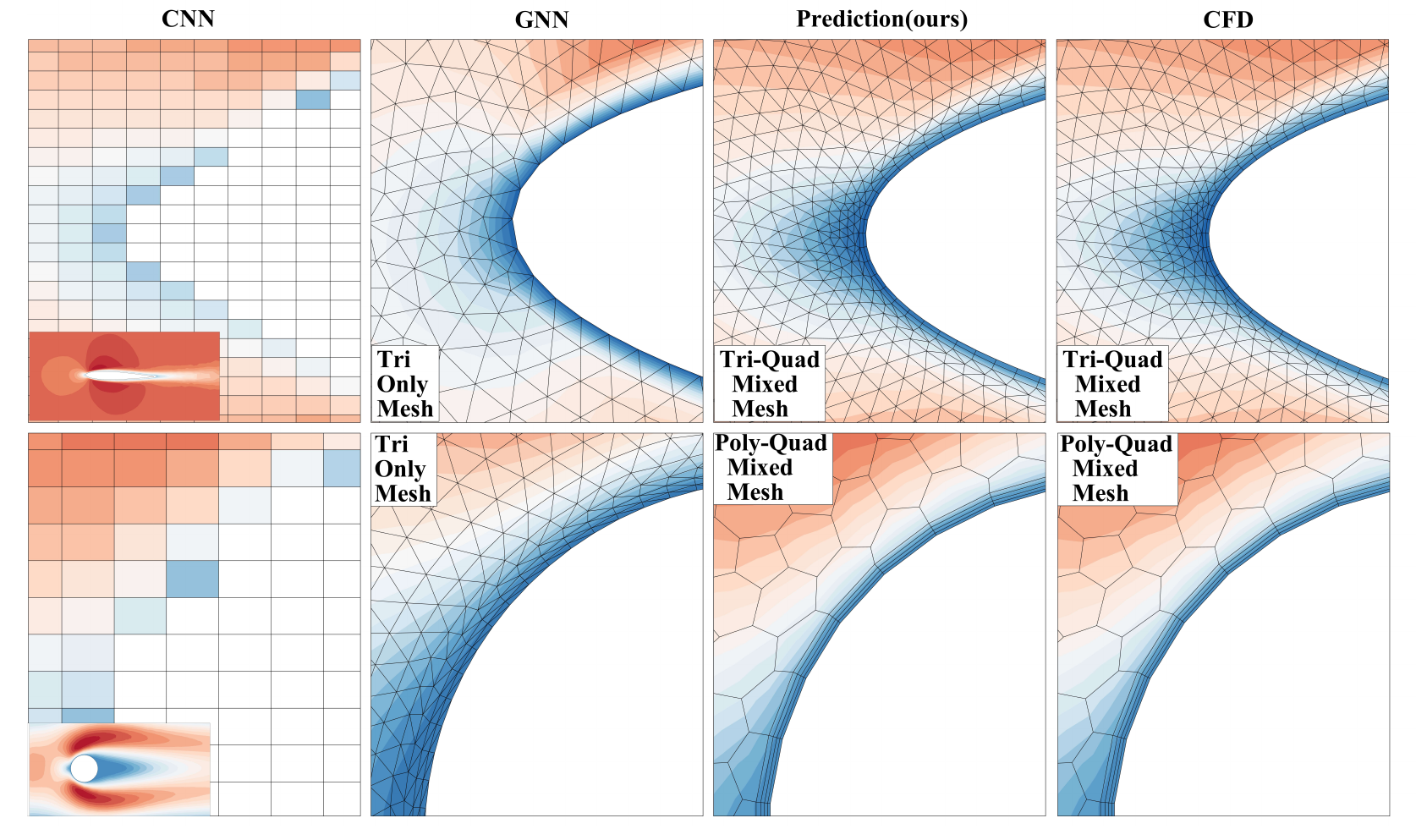}
    \end{minipage}
    \caption{Comparison of velocity field prediction results using different neural networks under various flow field characterization methods. 1) The flow field cloud diagram solved based on the U-net network, 2) The flow field prediction results of the graph neural network under a fully triangular discretized unstructured mesh\cite{pfaff2020learning,li2024predicting}, 3) The results of the unstructured mesh prediction under a mixed discretization of triangular and quadrilateral boundary layer meshes, 4) The results of CFD simulation.}
    \label{meshcompare}
\end{figure}

\begin{table}[!htbp]
\centering
\caption{Comparison of Different Neural Networks and Loss Function Performance, and Suitable Grid Types}
\scalebox{0.8}
{
\begin{tabular}{c|cccc}
\hline
\begin{tabular}[c]{@{}c@{}}Loss Type\\ Grid Type\\ Neural Network\end{tabular}  & \begin{tabular}[c]{@{}c@{}}Pre-computed \\ Data\end{tabular} & \begin{tabular}[c]{@{}c@{}}Solving New\\ Boundary \\ Condition \\ Without \\ Retraining\end{tabular} & \begin{tabular}[c]{@{}c@{}}Solving New\\ Source\\ Without\\ Retraining\end{tabular} & \begin{tabular}[c]{@{}c@{}}Geometry \\ Adaptability\end{tabular} \\ \hline

\begin{tabular}[c]{@{}c@{}}DD-Uniform Grid \cite{li2020fourier,wang2023swin}\\ CNN\end{tabular}          & \checkmark         & \checkmark   & \checkmark   & \redcircle         \\ \cline{1-1}

\begin{tabular}[c]{@{}c@{}}DD-Structured Grid\\ CNN \cite{chen_towards_2021,chen2024deep}\end{tabular}           & \checkmark         & \checkmark   & \checkmark   & \greencircle   \\ \cline{1-1}

\begin{tabular}[c]{@{}c@{}}DD-Unstructured Grid\\ GNN \cite{pfaff2020learning,fortunato_multiscale_2022,sanchez2020learning,seo2019physics}\end{tabular}         & \checkmark          & \checkmark   & \checkmark   & \greencircle   \\ \cline{1-1}

\begin{tabular}[c]{@{}c@{}}ND(FDM)-Uniform Grid\\ CNN \cite{wandel_teaching_2021,wandel2020learning,ranade_discretizationnet_2020}\end{tabular}   & \checkmark  & \ding{53}    & \ding{53}    & \redcircle         \\ \cline{1-1}

\begin{tabular}[c]{@{}c@{}}ND(FEM)-Unstructured Grid \\\ GNN \cite{gao_physics-informed_2022}\end{tabular}       & \ding{53}   & \ding{53}           & \ding{53}           & \greencircle   \\ \cline{1-1}

\begin{tabular}[c]{@{}c@{}}AD-Mesh Free\\ PINN \\ (baseline) \cite{rao_physics-informed_2020,raissi2019physics,jin_nsfnets_2021}\end{tabular}        & \ding{53}   & \ding{53}         & \ding{53}         & \greencircle   \\ \hline

\begin{tabular}[c]{@{}c@{}}ND(FVM)-Unstructured Grid\\ GNN(Ours)\end{tabular}            & \checkmark       & \checkmark             & \checkmark                & \greencircle                        \\ \hline

\end{tabular}
}
  \caption*{\textbf{*}:
             \greencircle \quad good ; 
             \redcircle \quad worse ; \\
            \quad \quad  DD - Data Driven ;
            \quad  ND - Numerical Differential ;
            \quad  AD - Auto Differential ;
            }
   \label{NN-grid-train compare}
\end{table}

However, these methods have certain limitations. For example, it is necessary to map to an equidistant MAC grid\cite{wandel2020learning}\cite{wandel_teaching_2021}\cite{ranade_discretizationnet_2020} to use the finite difference method, or to use structured grid combined with physical-to-computational domain coordinate system transformation methods\cite{chen_towards_2021} to adapt Convolutional Neural Networks (CNNs). For a uniformly pixelized grid, the flow field boundaries often produce large errors (as shown in Fig.\ref{meshcompare}), and for the coordinate transformation method\cite{chen_towards_2021}, applying the model to different grids faces issues with insufficient geometric shape generalization, and generating structured grids for complex geometries is a great challenge. For finite element methods\cite{gao_physics-informed_2022} or spline interpolation methods\cite{wandel_spline-pinn_2022}, although they can be performed on unstructured grids, such methods typically consume more computational resources because they use the automatic differentiation mechanism in the deep learning framework to solve physical-related derivatives. In summary, as can be seen from Tab.\ref{NN-grid-train compare}, we can conclude the following points:

\quad  1. Data-driven methods (including other neural operator methods not listed in the table, \cite{lu2021learning,li2023fourier,brandstetter2022message}) possess the capability to solve multiple boundary conditions and multiple source terms simultaneously using the same model. Furthermore, according to Ref.\cite{brandstetter2022message}, it is possible to solve multiple partial differential equations (PDEs) simultaneously within the same model training process. However, it cannot be denied that in the field of CFD, acquiring high-precision datasets is extremely challenging.

\quad  2. Current methods based on automatic differentiation and the application of soft boundary conditions \cite{raissi2019physics,zhu2019physics} typically can only solve one boundary condition or one source term in a single training session.

\quad  3. Pixel-style grids exhibit poor geometric adaptability(Fig.\ref{meshcompare}), which results in reduced boundary layer prediction accuracy.

Therefore, there is an urgent need to develop methods that can directly work on unstructured grids and use numerical differentiation to approximate the derivatives of physical equations, thereby introducing physical inductive bias to reduce dependency on high-precision datasets.

Graph Neural Networks (GNNs) are a type of neural network capable of processing data in non-Euclidean spaces\cite{sanchez2020learning,pfaff2020learning,seo2019physics,horie_physics-embedded_2023}. GNNs take graph structures composed of nodes and edges as input and learn the feature representations of nodes and edges through information propagation and aggregation\cite{gilmer_neural_nodate,battaglia2018relational,zhou_graph_2020}. The adaptability and scalability of GNNs have led to their widespread application in various fields, including social network analysis, recommendation systems, and knowledge graphs. Consequently, in recent years, researchers have proposed various types of GNNs to model the physical world, particularly in PDE solvers using GNNs \cite{li2024predicting,brandstetter2022message,horie_physics-embedded_2023,li_graph_2022,gao_physics-informed_2022,horie_physics-embedded_2023}.

The Finite Volume Method (FVM), a staple in Computational Fluid Dynamics (CFD), excels in handling complex geometries and multiphysics problems on unstructured grids. It discretizes physical domains, approximates fluxes at control volume boundaries, and converts partial differential equations into algebraic equations for iterative solving. 
Therefore, recently, many algorithms inspired by the finite volume method have emerged\cite{praditia_finite_2021, karlbauer_composing_2022,li2024predicting,ranade_discretizationnet_2020,jeon2022finite}. However, these algorithms have not been fully implemented on unstructured grids in terms of the discretization process of the finite volume method\cite{ranade_discretizationnet_2020} and, to varying extents, still require pre-computed data\cite{praditia_finite_2021, karlbauer_composing_2022,li2024predicting,jeon2022finite}. Therefore, Our goal is to combine the powerful non-Euclidean data representation ability of graph neural networks and finite volume methods \cite{aulakh_generalized_2022}\cite{ranade_discretizationnet_2020}\cite{fortunato_multiscale_2022}\cite{allen_physical_2022}, trying to simulate incompressible fluid flow in unstructured grid discretization. In this work, we propose a novel GNN-based PDE solver framework inspired by the finite volume method, which solves partial differential equation in a unified manner without the need for supervised data. The key points of this work can be summarized as follows:

\begin{itemize}
\item  We have implemented a differentiable finite volume method algorithm using deep learning frameworks, which allows us to construct FVM Loss for GNNs without the need for pre-computed data. As a result, Our model can simultaneously solve multiple boundary conditions, multiple source terms, and even multiple types of PDEs in a single training sessions.
\item  With the support of the finite volume method, we introduce a special variable arrangement that differs from traditional FVM, which doesn't require ghost cells to apply boundary conditions. Our method is very simple and accurately meets boundary conditions.
\item  We have designed a completely graph-based data structure, allowing our method to fully run various algorithms based on FVM in parallel on the GPU.
\end{itemize}

\section{Method}\label{sec:Method}

This study mainly focuses on the two-dimensional incompressible Navier-Stokes equations and Possion's equations, as shown below. Here, $t$ is the dimensionless time, Eq.\eqref{momtem} is the momentum equation, $\mathbf{u}(\mathbf{x},t)=[u,v]$ represents the velocity vector in the $x,y$ directions, $p$ is the pressure, $\rho$ is the fluid density, $\mu$ is the dynamic viscosity of the fluid, Eq.\eqref{continus_difference} is the continuity equation, $Re=\frac{U_{ref}D_{def}}{\mu}$ is the Reynolds number defined by characteristic length $D_{def}$, reference speed $U_{ref}$ and kinematic viscosity ${\mu}$, $\mathbf{f}$ is the volume force such as gravity, which is ignored in this study. In the subsequent discussions, $\Gamma_{D}$ and $\Gamma_{N}$ represent the Dirichlet and Neumann boundary conditions, respectively.

\begin{equation}\label{momtem}
\rho \frac{\partial \mathbf{u} }{\partial t} +\rho u\cdot \bigtriangledown \mathbf{u}=
-\bigtriangledown p+
\mu \bigtriangledown ^{2}\mathbf{u}+\rho\mathbf{f} \qquad in \quad \Omega,
\end{equation}

\begin{equation}\label{continus_difference}
\bigtriangledown \cdot \mathbf{u} =0 \qquad in \quad \Omega,
\end{equation}

\begin{equation}\label{eq2}
\mathbf{u}=\mathbf{u}_{\Gamma } \qquad on \quad \Gamma_{D},
\end{equation}

\begin{equation}\label{eq3}
\nu(\mathbf{n} \cdot \nabla) \mathbf{u}_{\Gamma} -p_{\Gamma}  \mathbf{n} =0 \qquad on \quad \Gamma_{N}
\end{equation}

As can be seen, both Eq.\eqref{momtem} and Eq.\eqref{continus_difference} have differential operators $ \bigtriangledown $, 
and it is relatively difficult to solve for the derivative on unstructured grids. 
Therefore, we need to use the finite volume method to convert the differential form of Eq.\eqref{momtem} and Eq.\eqref{continus_difference} into an integral form.

Generally, the finite volume method is used to solve the integral form of NS equations satisfying conservation laws,
applying it to Eq.\eqref{momtem} and Eq.\eqref{continus_difference} can derive \cite{moukalled2016finite} Eq.\eqref{NS-integration-continus}, 
Eq.\eqref{NS-integration-x-mom} and Eq.\eqref{NS-integration-y-mom}. We have converted the differential form of the NS equation into an integral of flux over grid cells (control volumes).

\begin{equation}\label{NS-integration-continus}
\int_{V}^{} \bigtriangledown \cdot \mathbf{u} dV=\int_{S} \mathbf{u} \cdot \mathbf{n} \mathrm{d} S=0
\end{equation}
\begin{equation}\label{NS-integration-x-mom}
\frac{\partial}{\partial t} \int_{V} u \mathrm{~d} V=-\oint_{S} u \mathbf{u} \cdot \mathbf{n} \mathrm{d} S-\frac{1}{\rho} \oint_{S} p \cdot n_{x} \mathrm{~d} S+\mu \oint_{S} \nabla u \cdot \mathbf{n} \mathrm{d} S
\end{equation}
\begin{equation}\label{NS-integration-y-mom}
\frac{\partial}{\partial t} \int_{V} v \mathrm{~d} V=-\oint_{S} v \mathbf{u} \cdot \mathbf{n} \mathrm{d} S-\frac{1}{\rho} \oint_{S} p \cdot n_{y} \mathrm{~d} S+\mu \oint_{S} \nabla v \cdot \mathbf{n} \mathrm{d} S
\end{equation}

The above equations are the integral form of the NS equations for 2D incompressible flow. Eq.\eqref{NS-integration-continus} represents the continuity equation, Eq.\eqref{NS-integration-x-mom} represents the momentum equation in the $x$ direction, and Eq.\eqref{NS-integration-y-mom} represents the momentum equation in the $y$ direction. $\int_{V}$ is a volume integral and $\oint_{S}$ is a surface integral. 

\subsection{Least Square Method for Gradient Reconstruction}
Gradient reconstruction is the core process in solving  Eq.\eqref{NS-integration-continus}, Eq.\eqref{NS-integration-x-mom} and Eq.\eqref{NS-integration-y-mom}. On unstructured grids, gradient reconstruction is commonly divided into two methods: the Green-Gauss method and the WLSQ (Weighted Least Squares) method. Given the complexity of the meshes in the subsequent examples, this paper primarily uses the WLSQ method to solve for the gradients of physical variables.

The use of least squares to estimate the gradient of the cell center was initially proposed by \cite{barth19913,barth1992aspects}, which can be understood as a $k$-exact\cite{hui2019accuracy} ($k=1$) reconstruction in the current element and its neighborhood. It assumes a linear function distribution that satisfies not only the average value constraint of the current element but also the average value constraint of adjacent elements. In the case of two dimensions, the current element is denoted as 0, and the adjacent elements are denoted as $i (i=1, 2, ..., N)$. This leads to the following least squares problem.
\begin{equation}
\label{lsq}
\centering
\left[\begin{array}{cc}
w_{1} \Delta x_{1} & w_{1} \Delta y_{1} \\
w_{2} \Delta x_{2} & w_{2} \Delta y_{2} \\
\vdots & \vdots \\
w_{i} \Delta x_{i} & w_{i} \Delta y_{i} \\
\vdots & \vdots \\
w_{N} \Delta x_{N} & w_{N} \Delta y_{N}
\end{array}\right]\left[\begin{array}{c}
\frac{\partial \phi}{\partial x}  \\
\frac{\partial \phi}{\partial y} 
\end{array}\right]=\left[\begin{array}{c}
w_{1}\left(\phi_{1}-\phi_{0}\right) \\
w_{2}\left(\phi_{2}-\phi_{0}\right) \\
\vdots \\
w_{i}\left(\phi_{i}-\phi_{0}\right) \\
\vdots \\
w_{N}\left(\phi_{N}-\phi_{0}\right)
\end{array}\right]
\end{equation}
Among them, $\omega_i=\frac{1}{\sqrt{\Delta x_{F_{k}}^{2}+\Delta y_{F_{k}}^{2}}}$ is the weight function, which can be taken as a function of the distance from the centroid of element $i$ to the centroid of element $0$. The least squares problem can be represented as $Ax=b$. In order to avoid ill-posed problems, the $A$ matrix of the aforementioned least squares problem is decomposed into the product of an orthogonal matrix $Q$ and an upper triangular matrix $R$ using the Gram-Schmidt process\cite{winkler2012reduced}. Therefore, we obtain $x = R^{-1}Q^Tb$. Singular value decomposition can also be used to estimate the gradient of the cell center by using the generalized inverse of matrix $A$.


\subsection{Generative (Unsupervised) Finite Volume Graph Network: Gen-FVGN}\label{sec:Gen-FVGN}
In this subsection, we introduced the complete finite volume method discretization solution process into FVGN and named it Gen-FVGN, which stands for Generative or Unsupervised FVGN. Gen-FVGN can provide solutions to the corresponding PDEs or generate physically consistent flow fields using only the grid, boundary conditions, and specified equation types. The general forward and training pool architecture of Gen-FVGN is illustrated in Fig.\ref{fig:Gen-FVGN-forward}. The overall architecture of Gen-FVGN can be broadly divided into three parts: the backbone neural network, the construction of PDE loss, and the training logic.

\subsubsection{Neural Network Architecture}\label{sec:Gen-FVGN-arc}
The neural network architecture of Gen-FVGN is largely inherited from FVGN (as described in Ref.\cite{li2024predicting}). It follows an Encoder-Processor-Decoder structure, where the Processor also uses a two-step message aggregation GN block with extended templates from FVGN. There are two main differences between Gen-FVGN and FVGN in terms of network architecture. First, Gen-FVGN positions physical quantities such as $u, v, p$ at the vertices of the grid, whereas FVGN places them at the cell centers. Thus, FVGN can be considered Cell-Centered, while Gen-FVGN is Vertex-Centered. There are two main reasons for this approach: 1), it facilitates the imposition of boundary conditions (as discussed in Sec.\ref{sec:hard-enforce-boundary}); 2), Gen-FVGN directly decodes the physical quantities at the grid vertices for the next time step or iteration, unlike FVGN, which decodes the flux of the right-hand side terms of the Navier-Stokes equations at the face elements. Below is a more detailed description of the architecture.

\begin{figure}[!htbp]
\centering
\begin{minipage}{\linewidth}\centering
    \includegraphics[width=1\textwidth]{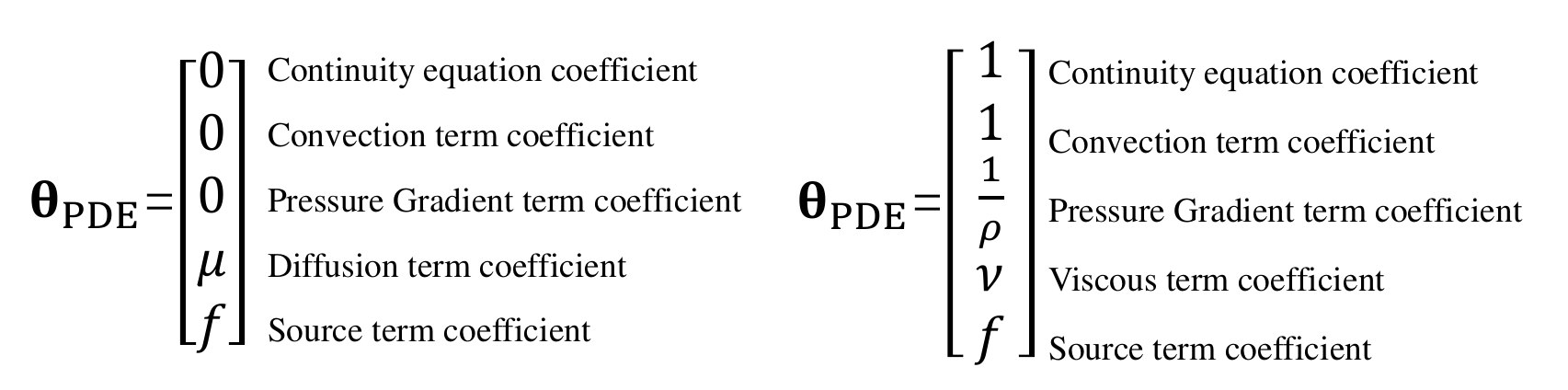}
    \end{minipage}
    \caption{Left: When solving Poisson's equation, Continuity equation coefficient, Convection term coefficient, and Pressure Gradient term coefficient are set 0. Right: When solving Navier-Strokes equation, all the coefficients are maintained for the dimensionless NS equation.}
    \label{fig:Theta_pde}
\end{figure}

\noindent$\textbf{Encoder}$ \quad The encoder consists of two parts: the vertex attribute encoder $\phi^{v}$ and the edge attribute encoder $\phi^{e}$. The encoder only encodes the current flow field centered on the vertex, $G_v(V^{t}, E_v)$, into a directed graph $\mathcal{G}_{v}(\mathbf{v}^{\prime}, \mathbf{e}_{v_{ij}}^{\prime})$, where each vertex and mesh edge possesses hidden layer features $\mathbf{v}^{\prime}$ with a size of 128. The edge features have the same dimensions as the cell center. Additionally, the input at the vertex for $\phi^{v}$ includes the velocity fields in the $x$ and $y$ directions and the pressure field at time step $t$, $\mathbf{v}^{t}=[u,v,p]$, a one-hot encoding vector $\mathbf{n}$ for the specified node type (boundary node or internal node), and most importantly $\theta_{pde}$. Here $\theta_{pde}$ represents the specific equation currently being solved, as shown in Fig.\ref{fig:Theta_pde}, which is a 5-dimensional vector that encodes the individual terms' coefficients of the equation being solved. This allows the neural network to differentiate the distribution patterns of solutions between different equations. The input at the edge for $\phi^{e}$ includes the differences of adjacent cell units at each edge \cite{brandstetter2022message,seo2019physics}, edge length, and the difference in relative coordinates forming each edge, which can be represented as $\mathbf{e}_{v_{ij}}^{t} = \phi^{e}(\mathbf{u}_{i}-\mathbf{u}_{j},\mathbf{x}_{i}-\mathbf{x}_{j},\left | \mathbf{x}_{i}-\mathbf{x}_{j} \right |)$. Both $\phi^{v}$ and $\phi^{e}$ are implemented using two-layer linear layer MLPs with SiLU \cite{elfwing2018sigmoid} activation functions and output after a LayerNorm layer.

\noindent$\textbf{Processor}$ \quad The processing sequence of Gen-FVGN consists of \( M \) identical message passing layers, which encompass the Graph Network blocks \cite{sanchez2018graph}. However, there are some differences from the message passing process in FVGN. In FVGN's Cell-Block, edge features that share a point are first aggregated to the vertex, and then the features of vertices that constitute a cell are aggregated to the cell center. Since Gen-FVGN is Vertex-Centered, its Node Block first aggregates the edge features that make up the same cell to the cell center, a process referred to as the first message aggregation. Subsequently, the cell features that share a vertex are aggregated to that vertex, termed the second message aggregation. The original vertex features are then concatenated and updated through a \( Node-MLP \). Similarly, this MLPs have two hidden layers of size 128 and are composed of SiLU activation functions and Layernorm. The Edge Block of Gen-FVGN concatenates the features of the two vertices forming an edge to the current edge feature, which is then updated through an \( Edge-MLP \) with the same architecture as the \( Node-MLP \). Specifically, the processes within the Node Block can be expressed by the first two equations in \eqref{Gen-FVGN-processor}, while the process in the Edge Block can be represented by the last equation in Eq.\eqref{Gen-FVGN-processor}.

\begin{figure}[!htbp]
\centering
\begin{minipage}{\linewidth}\centering
    \includegraphics[width=0.9\textwidth]{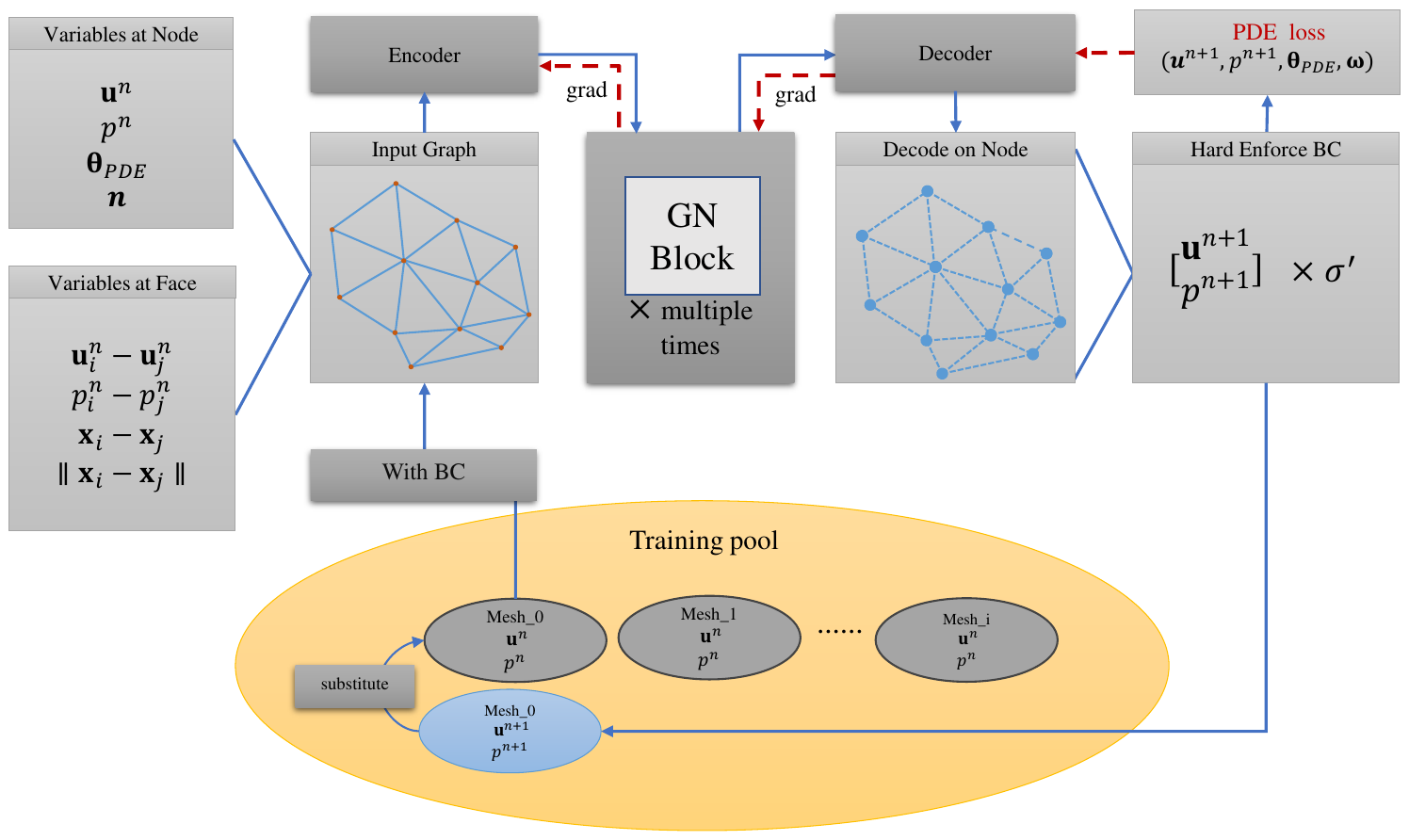}
    \end{minipage}
    \caption{Generative Physics Encode-FVGN forward process. There's only PDE equation loss without boundary loss during training process. And $\sigma ^{'}$ represents channel mask, e.g. when input graph is 2D Poisson's equation $\sigma ^{'}=\left [ 1,0,0 \right ]$, and when input graph is 2D Navier-Strokes equation $\sigma ^{'}=\left [ 1,1,1 \right ]$, because $\left [ u, v, p\right]$ is only need for NS equation. $\omega$ is the learnable parameters inside the neural network. Superscript $n$ represents iteration times, $\mathbf{n}$ stands for the one-hot node type vector, e.g. the interior node will set to be 0, and wall node are 6 \ldots etc.}
    \label{fig:Gen-FVGN-forward}
\end{figure}

\begin{equation}\label{Gen-FVGN-processor}
\overline{\mathbf{c}}_{i}^{\prime} \leftarrow \dfrac{1}{N_c} \left(\sum_{e \in cell_i} {}  \mathbf{e}_{v_{ij}}^{\prime} \right) ;\quad \overline{\mathbf{v}}_{i}^{\prime} \leftarrow \phi^{vp}\left(\mathbf{v}^{\prime}_{i}, \dfrac{1}{N_v} \sum_{}^{} \overline{\mathbf{c}}_{i}^{\prime},\right) ;\quad  \overline{\mathbf{e}}_{v_{ij}}^{\prime} \leftarrow \phi^{ep}\left(\mathbf{e}_{v_{ij}}^{\prime},\overline{\mathbf{v}}^{\prime}_{i}, \overline{\mathbf{v}}^{\prime}_{j}\right) 
\end{equation}

In Eq.\eqref{Gen-FVGN-processor}, \( N_c \) represents the number of edges constituting a cell, and \( N_v \) indicates the number of cells sharing the same vertex. Here, the "two-step message aggregation" process in Gen-FVGN slightly differs from that in FVGN because the former places variables at the grid vertices. Therefore, in the message aggregation process, messages should be aggregated to the cell's vertices in the final step. This adjustment ensures that the variable placement aligns with the Vertex-Centered approach of Gen-FVGN, facilitating the incorporation of boundary conditions and the direct update of physical quantities at the vertices.

\noindent$\textbf{Decoder}$ \quad The decoder of Gen-FVGN employs the same $MLP$ architecture as FVGN, composed of 2 hidden layers with 128 hidden units each and SiLU activation functions, designated as $\phi^{d}$. Unlike the previously mentioned $MLP$s, $\phi^{d}$ does not incorporate a LayerNormalization (LN) layer. The decoder's task is to decode the latent vectors at the vertices after the message passing layer into physical quantities that adhere to the laws of physics or, more specifically, to the solutions of the PDEs. For instance, for the Navier-Stokes equations or the wave equation, the decoder outputs the values of $u, v, p$ for the next iteration or time step, while for the Poisson equation, it only outputs $u$.

To standardize the decoding process for these three equations or to simultaneously solve these three equations in the same training cycle, the output of $\phi^{d}$ is still set as $\mathbf{v} \in \mathbb{R}^{n \times 3}$, but we introduce a multiplier $\sigma^{\prime}$ (as shown in Fig.\ref{fig:Gen-FVGN-forward}). The role of $\sigma^{\prime}$ is to adjust the output vector according to the type of equation being solved: for a two-dimensional Poisson equation, $\sigma^{\prime} = \left[ 1,0,0 \right]$, and for a two-dimensional Navier-Stokes equation, $\sigma^{\prime} = \left[ 1,1,1 \right]$. Thus, the final output of Gen-FVGN is effectively $\mathbf{v} \times \sigma^{\prime}$. This method ensures that the model can handle different types of PDEs while maintaining a unified output format.

\subsubsection{Training Process}\label{sec:Gen-FVGN-train-strategy}
Similar to FVGN, the training of Gen-FVGN still adopts the One-Step training approach. That is, we use the flow field from the previous round, $\mathbf{u}^{tn}$, as input and then predict the flow field for the next iteration or the next time step, $\mathbf{u^{n+1}}$. However, the difference lies in the fact that at the initial moment $t=0$, the input for Gen-FVGN is entirely determined by boundary conditions and is a uniform field. As training commences, Gen-FVGN incorporates each round’s output back into the original flow field. In other words, once training begins, Gen-FVGN maintains a training pool $\mathfrak{G}$, as illustrated in the lower part of Figure \ref{fig:Gen-FVGN-forward}. This pool contains different grids $G_{1}^{n},\quad G_{2}^{n},\quad G_{3}^{n},\quad...$. Gen-FVGN then draws a mini-batch of $\mathcal{G}^{n}$ from $\mathfrak{G}$ to send into the GN block for message passing, followed by decoding through a decoder to predict the flow field of the next iteration or time step, $\mathcal{G}^{n+1}$. After computing the PDE Loss, it is returned to $\mathfrak{G}$ to await the next round of extraction. Ultimately, since the size of $\mathfrak{G}$ is fixed, the number of boundary conditions learned in one training round is limited. Therefore, after each iteration, we randomly reset one of the flow fields in $\mathfrak{G}$, $G_{k}^{n}$, to $G_{k}^{0}$, and in the process of resetting, we also reselect the boundary conditions or source terms. This approach enables a broader range of learning for boundary conditions and source terms.

\subsection{Graph Representation \& Variable Arrangement}\label{sec:Graph-rep}
\begin{figure}[!htbp]
\centering
\begin{minipage}{\linewidth}\centering
    \includegraphics[width=1\textwidth]{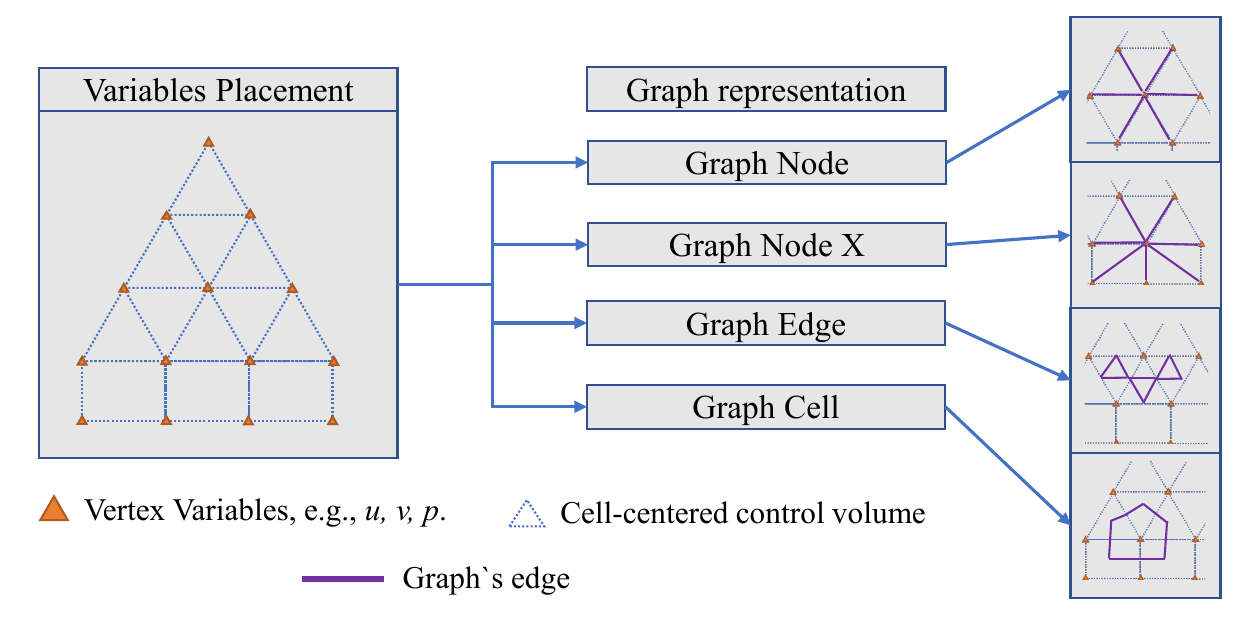}
    \end{minipage}
    \caption{Graph Representation \& Variable Arrangement}
    \label{fig:Graph_rep}
\end{figure}

Similar to what was done in Ref.\cite{li2024predicting}, unstructured grids can be represented as multiple graph data structures. As shown in Fig.\ref{fig:Graph_rep}, we can use different mesh entities as graph vertices and construct the graph data structure based on the adjacency relationships between these entities. For example, using mesh vertices as graph vertices and mesh edges as adjacency relations can construct a Graph Node. More details on the data structure can be found in Sec.\ref{sec:Gen-FVGN-datastructure}.

It must be noted here that the variable arrangement method used in this paper is different from the traditional FVM. This paper arranges physical variables at the mesh vertices, but when calculating the FVM Loss, a Cell-Centered control volume is used (as shown in Fig.\ref{fig:Graph_rep}), i.e., the input to the vertex encoder and the output from the vertex decoder are vertex-wised, while the FVM Loss is cell-wised. This approach brings the following four advantages:

\quad 1. A simple method of imposing boundary conditions (details can be seen in Sec.\ref{sec:hard-enforce-boundary})

\quad 2. Lower computational overhead on triangular meshes since the number of vertices is much less than the number of cells \cite{li2024predicting}

\quad 3. Compared to the traditional FVM which places both the original variables and the variables to be solved at the vertices or cells, the method used in this paper has more stable gradient reconstruction accuracy on various types of unstructured grids (details in Sec.\ref{sec:grad-verify})

\quad 4. Almost no need for non-orthogonality error correction. In unstructured grids, the midpoint of the line connecting cells may not coincide with the midpoint of the shared edge, hence when interpolating cell gradients to the face, non-orthogonality correction is required\cite{moukalled2016finite}. However, the method used in this paper does not require this process.

\subsection{Hard Enforcement of Boundary Conditions}\label{sec:hard-enforce-boundary}
\begin{figure}[!htbp]
\centering
\begin{minipage}{\linewidth}\centering
    \includegraphics[width=1\textwidth]{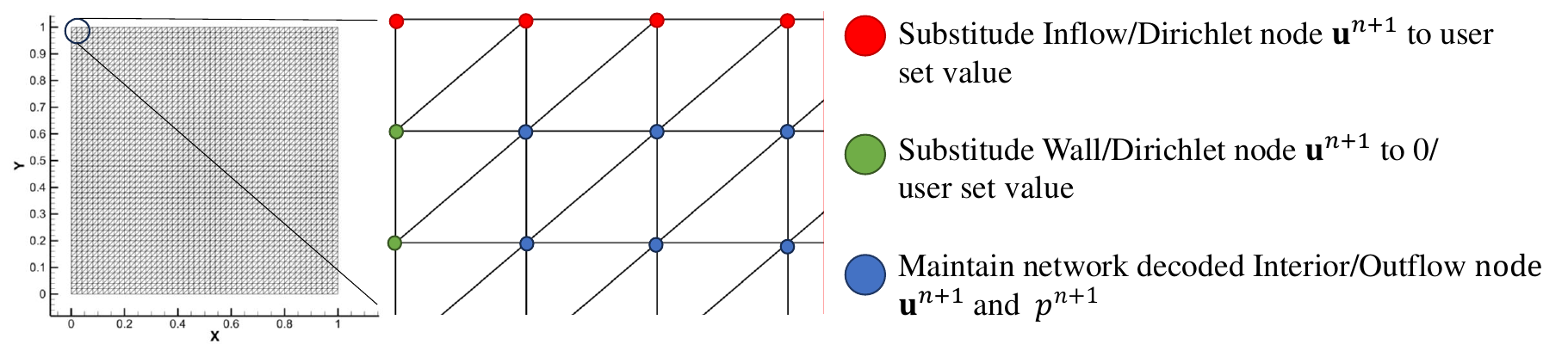}
    \end{minipage}
    \caption{By hard enforcing boundary conditions, Gen-FVGN can always maintain high accuracy prediction near the boundary cells, Because there's absolutely no "Boundary Loss", compared to traditional PINN method(or soft/penalty method)}
    \label{Hard_Enforce_BC}
\end{figure}

\noindent$\textbf{Dirichlet boundary condition}$ \quad We follow a similar way in hard enforcement of boundary conditions by Han Gao\cite{gao_phygeonet_2021}. But there's also a tiny difference, in the conventional FVM or FDM, a ghost cell method was usually employed alongside the boundary cell and all variables are placed in the cell center. While we placed all decoded variables at the vertex, so we don't need a ghost cell to enforce Dirichlet BCs. We processed it simply, which substituted the decoded vertex value of the boundary node with true boundary conditions(see Fig.\ref{Hard_Enforce_BC}). So the whole decoded field output by the neural network will satisfy Dirichlet BCs automatically. 

\noindent$\textbf{Neuman boundary condition}$ \quad Eq.\eqref{pressure outlet} describes how we apply pressure boundary conditions in the Navier-Stokes (NS) equations. Here, $p_0$ represents the outlet pressure set by the user. To prevent backflow, we thus have $\hat{p_{0}} \leq p_{0}$, indicating that the predicted pressure at the outlet boundary should be less than the pressure specified by the user. Therefore, by simply rearranging the terms, we can derive the loss function at the outlet (Eq.\eqref{loss_pressure_outlet}). The reason we set the pressure outlet as a loss function is because, for incompressible problems, the pressure boundary is crucial for global solution accuracy. We have integrated methods used in various commercial software, hence we decided to set the outlet pressure equal to the viscous forces.

\begin{equation}\label{pressure outlet}
{[-p \mathbf{I}+\mu \bigtriangledown \mathbf{u} ] \mathbf{n}=-\hat{p_{0}} \mathbf{n}} , \quad
\hat{p_{0}} \leq p_{0} \quad \text { on } \partial \Omega^{N} 
\end{equation}

\begin{equation}\label{loss_pressure_outlet}
L_p=\left \|   [-p \mathbf{I}+ \mu \bigtriangledown \mathbf{u} ] \mathbf{n} +\hat{p_{0}} \mathbf{n}\right \|_2
\end{equation}


\subsubsection{Essential FVM implementation requirements}
Our designed set of five index variables encompasses Cells node, Cells face, Cells index, Edge index, and Neighbor Cell. Specifically, the Cells node delineates which vertices constitute a cell, while the Cells face defines which edges/faces compose the current cell. Cells index maintains identical dimensions and structures as Cells node and Cells face, thereby establishing the relations between vertices/edges/faces and cells. Moreover, the Edge index merely identifies which two points form an edge, and Neighbor Cell consists of the indices of cells adjacent to the current edge/face. A concrete example is provided, as depicted in the lower section of Fig.\ref{sec:Gen-FVGN-datastructure}. Utilizing a differentiable function, Scatter, we compute and transfer the values from the face centers to the cell centers, a quintessential integration operation in finite volume methods. In practical implementations, face values are substituted with physical variables such as unit normal vectors $u, v, p$, and face areas.

\subsection{Finite Volume Informed Loss Function Construction}
As previously mentioned, following the output $\mathbf{v}$ of Gen-FVGN, the process of constructing the PDE loss begins, as shown in Fig.\ref{fig:Gen-FVGN-forward}. Here, we use the construction process of the Navier-Stokes (NS) equations as an example. From Eq.\eqref{momtem}, it can be seen that the right-hand side terms of the momentum equation mainly include the time derivative term $\frac{\partial \mathbf{u}}{\partial t}$, the convection term $(\mathbf{u} \cdot \bigtriangledown)$, the pressure term $\bigtriangledown p$, and the viscosity term $\bigtriangledown \cdot \bigtriangledown \mathbf{u}$. After integrating these over the volume and applying the divergence theorem, they can be converted into surface integral form. Obviously, our variable arrangement allows us to interpolate various physical quantities to the faces with high accuracy. When constructing the Loss, we did not strictly follow the traditional Vertex-Centered control volume layout. Our control volumes are still Cell-Centered. In short, we decode the physical variables at the vertices, but the constructed PDE equations are located at the cell. This approach brings a significant advantage: the hard imposition of boundary conditions. This is different from the traditional FVM, which needs to construct Ghost Cells to impose boundary conditions \cite{moukalled2016finite,gao_phygeonet_2021,sun2020surrogate}, and also different from the traditional PINN's method of softly imposing BC loss. Specific details can be found in Sec.\ref{sec:hard-enforce-boundary}. Here, we will introduce the specific discretization process item by item.

\subsubsection{Time Derivative Term}
For the time derivative term, we use a simple explicit-implicit format, as employed by Ref.\cite{wandel2020learning}. Discretization in the time domain is a necessary step in handling the time derivative of the velocity field $\frac{\partial \mathbf{u}}{\partial t}$. In Eq.\eqref{momtem}, it can be rewritten as:
\begin{equation}\label{eq:imex-momtem}
    \rho\left(\dfrac{\mathbf{u}^{t+d t}-\mathbf{u}^{t}}{d t}+\left(\mathbf{u}^{t^{\prime}} \cdot \nabla\right) \mathbf{u}^{t^{\prime}}\right)=-\nabla p^{t+d t}+\mu \Delta \mathbf{u}^{t^{\prime}}+\vec{f}
\end{equation}
The objective is to maintain stability and accurate solutions while using as large a time step $dt$ as possible. The stability and accuracy of the NS equation solving process largely depend on the definition of $\mathbf{u}^{t'}$. Choosing $\mathbf{u}^{t'} = \mathbf{u}^t$, commonly known as the explicit integration method, is strictly constrained by the CFL number, which greatly limits the choice of mesh size and time step, and often leads to unstable behavior. Selecting $\mathbf{u}^{t'} = \mathbf{u}^{t+dt}$ is typically associated with implicit integration methods and provides stable solutions at the cost of numerical dissipation. The Implicit-Explicit (IMEX) method, setting $\mathbf{u}^{t'} = (\mathbf{u}^t + \mathbf{u}^{t+dt}) / 2$, represents a compromise between the two methods and is considered to be more accurate, though not as stable as the implicit method.

\subsubsection{Continuity Equation and Convection Term}
For the continuity equation and the convection term, both can be transformed using the divergence theorem. Specifically, the convection term in conservative form can be expressed as $\bigtriangledown \cdot (\mathbf{u} \mathbf{u})$. That is, after decoding the velocity $\mathbf{u}_{v}$ at the vertices, it can be directly averaged to the faces that are Cell-Centered. In this case, the non-orthogonality errors in the interpolation process are particularly small, because for the Cell-Centered control volumes, the centers of the faces are the geometric centers of the edges. This method avoids the non-orthogonality errors caused by placing variables at the cell center and then interpolating them to the faces, as in traditional FVM. However, it must be noted that in this discretization format, through our experiments, we have found that the equation convergence process is particularly slow, i.e., the neural network's learning process for the surface integrals is very slow. But if we directly use the volume integral form of the equations, it greatly enhances the convergence speed of the neural network.

This discretization format is based entirely on the non-conservative form of the Navier-Stokes (NS) equations to construct the PDE loss. As shown in Eq.\eqref{momtem}, the convection term and the continuity equation actually include a $\bigtriangledown \cdot$ operator, which can be written in component form in two dimensions as $\frac{\partial }{\partial x} + \frac{\partial }{\partial y}$. In fact, as long as we can reconstruct the gradient of the corresponding physical quantity
$\bigtriangledown = 
\begin{bmatrix}
\frac{\partial }{\partial x}, \frac{\partial }{\partial y}
\end{bmatrix}^{T}$,
then we can calculate both the continuity equation and the convection term. That is:

\begin{equation} \label{eq:Gen-FVGN-cont}
\int_{\partial \Omega }^{ } \bigtriangledown \cdot \mathbf{u} \mathrm{d}V = 0 \Rightarrow 
\bigtriangledown \cdot \mathbf{u}_c \bigtriangleup V_c =0 \Rightarrow 
\dfrac{\partial u}{\partial x} + \dfrac{\partial v}{\partial y}=0
\end{equation}

\begin{equation}\label{eq:Gen-FVGN-convection}
\int_{\partial \Omega  }^{} \mathbf{u} \cdot \bigtriangledown \mathbf{u}\mathrm{d}V = 
\mathbf{u}_c \cdot \bigtriangledown \mathbf{u}_c \bigtriangleup V\Rightarrow 
\begin{Bmatrix}
(u_c\dfrac{\partial u}{\partial x} +v_c \dfrac{\partial u}{\partial y})\bigtriangleup V \\
(u_c\dfrac{\partial v}{\partial x} +v_c \dfrac{\partial v}{\partial y})\bigtriangleup V
\end{Bmatrix}
\end{equation}

Here, Eq.\eqref{eq:Gen-FVGN-cont} represents the continuity equation, where $\mathbf{u}_c = [u_c, v_c]^{T}$ are the values at the cell center, and $\bigtriangleup V$ is the area of the cell in two dimensions. Since the neural network's decoding output is on the Grid vertices, Eq.\eqref{eq:inteplot-2nd} can be used to interpolate the decoded values from the vertices to the cell center. Eq.\eqref{eq:Gen-FVGN-convection} represents the discretization of the convection term and its component formula in two dimensions. For the conservative NS equations, the convection term can also be written as:
\begin{equation}\label{eq:Gen-FVGN-convection-conserved}
\int_{\partial \Omega  }^{} \bigtriangledown \cdot (\mathbf{u} \mathbf{u})\mathrm{d}V = 
\bigtriangledown \cdot (\mathbf{u} \mathbf{u}) \bigtriangleup V \Rightarrow 
\begin{Bmatrix}
(\dfrac{\partial u_cu_c}{\partial x_c} +\dfrac{\partial u_cv_c}{\partial y})\bigtriangleup V \\
(\dfrac{\partial v_cu_c}{\partial x_c} +\dfrac{\partial v_cv_c}{\partial y})\bigtriangleup V
\end{Bmatrix}
\end{equation}

From our practice, in the solution process of the incompressible Navier-Stokes equations, the results between Eq.\eqref{eq:Gen-FVGN-convection-conserved} and Eq.\eqref{eq:Gen-FVGN-convection} are not significantly different. However, if one chooses to use Eq.\eqref{eq:Gen-FVGN-convection-conserved} to construct the loss function, then an additional gradient reconstruction is required. For Eq.\eqref{eq:Gen-FVGN-convection}, the entire process of building the loss function only requires one gradient reconstruction. That is, after the neural network decodes the values at the vertices $[u_v, v_v, p_v]$, using a gradient reconstruction method, the gradients
$\bigtriangledown = 
\begin{bmatrix}
\frac{\partial }{\partial x}, \frac{\partial }{\partial y}
\end{bmatrix}^{T}$
are reconstructed and then interpolated to the cell center. Thus, the continuity equation, convection term, and viscous term (see next Sec.\ref{sec:Gen-FVGN-p-and-vis-term}) can all be constructed.

\subsubsection{Pressure Term and Viscous Term}\label{sec:Gen-FVGN-p-and-vis-term}
From the results of the gradient reconstruction in the previous section, we can easily obtain the velocity field gradient $\bigtriangledown \mathbf{u}$, and similarly, we can also obtain $\bigtriangledown p$. The volume integral of the pressure term can thus be written as $\int_{\Omega} \bigtriangledown p \mathrm{d}V = \bigtriangledown p_c \bigtriangleup V$, where $\bigtriangleup V$ is the cell volume, which in two dimensions is the cell area $\bigtriangleup S$.

For the viscous term, the process is even simpler. By reconstructing the gradient at the Grid vertices $\bigtriangledown \mathbf{u}$, and using arithmetic averaging to interpolate it to the faces, we similarly avoid the interpolation errors common in traditional methods. Therefore, we can write out the integral process for the viscous term: $\int_{\Omega} \bigtriangledown \cdot \bigtriangledown \mathbf{u} \mathrm{d}V = \sum \bigtriangledown \mathbf{u}_f \cdot \mathbf{n} \bigtriangleup S$, where in two dimensions $\bigtriangleup S$ is actually the length of the control volume boundary $\bigtriangleup l$.
Finally, we can construct the aforementioned terms using the Weighted Squared Least Squares (WSLQ) method, resulting in the PDE Loss as follows:

\begin{equation}\label{eq:Gen-FVGN-loss_cont}
       L_{continuity}=\left \| \bigtriangledown \cdot \mathbf{u}^{t+1} \bigtriangleup V \right \|_2
\end{equation}
\begin{align}
\label{eq:Gen-FVGN-loss_mom}
L_{momentum} = &\, \left \| \dfrac{\mathbf{u}_{\overline{c}}^{t+d t} -\mathbf{u}_{\overline{c}}^{t}}{d t} \bigtriangleup V +\left(\mathbf{u}_{\overline{c}}^{t^{\prime}} \cdot \nabla\right) \mathbf{u}_{\overline{c}}^{t^{\prime}}\bigtriangleup V \right. \\
&\left.+ \frac{1}{\rho } \nabla p_{\overline{c}}^{t+d t}\bigtriangleup V 
- \nu \sum_{f\in cell} \bigtriangledown  \mathbf{u}_{\overline{f} }^{t^{\prime}}\cdot \mathbf{n}\bigtriangleup S 
- \vec{f}\bigtriangleup  V \right \|_2
\end{align}

Here, $\overline{c}$ and $\overline{f}$ represent the interpolated values derived from the neural network decoded values at the vertices, interpolated using Eq.\eqref{eq:inteplot-2nd}.

\begin{algorithm}[!htbp]
\caption{Learning the FVM discretization Loss}
\label{alg:train_process}
\DontPrintSemicolon  
 \KwIn{$G_n(V_n,E_n),G_{nx}(V_{nx},E_{nx}),G_{e}(V_{e},E_{e}),G_{c}(V_{c},E_{c})$, $[\mathbf{u}_{n}^{t},p_n^t,\mathbf{\theta}_{pde}] \in V_n$...}

\KwOut{Converged $\mathcal{NN}$ parameters}

\SetKwInput{KwInput}{Input}  
\SetKwInput{KwOutput}{Output} 
\While{True}{
$[\mathbf{u}_{n}^{t+1},p_n^{t+1}] \times \sigma^{\prime}$ $\leftarrow$ $\mathcal{NN}(G_n,G_c)$\;

Let $\mathbf{u}_{n}^{t+1}=\mathbf{u}_{\Gamma} \quad \text{on} \quad \partial_{\Omega}$ \;

Let $\mathbf{N}=[\mathbf{u}_{n}^{t},p_n^t,\mathbf{u}_{n}^{t+1},p_n^{t+1}]$\;

$\bigtriangledown \mathbf{N}$ $\leftarrow$ WLSQ($G_{n}, G_{nx}, \mathbf{N}$)\;

$ [\mathbf{C}, \bigtriangledown \mathbf{C}] $ $\leftarrow$ Interpolate($G_{n}, G_{c}, \mathbf{N}, \bigtriangledown \mathbf{N}$)\;

$L_{continuity}=\bigtriangledown \cdot \mathbf{u}^{t+1} \bigtriangleup V$ $\leftarrow$ $\mathcal{I}_{cont} (G_c,\bigtriangledown \mathbf{C},\mathbf{\theta}_{pde})$\;

$L_{momentum} = \left\| \frac{\mathbf{u}_{\overline{c}}^{t+dt} - \mathbf{u}_{\overline{c}}^{t}}{dt} \bigtriangleup V + (\mathbf{u}_{\overline{c}}^{t'} \cdot \nabla) \mathbf{u}_{\overline{c}}^{t'}\bigtriangleup V \right. $\;
$\left. + \frac{1}{\rho} \nabla p_{\overline{c}}^{t+dt}\bigtriangleup V - \nu \sum_{f \in \text{cell}} \bigtriangledown \mathbf{u}_{\overline{f}}^{t'} \cdot \mathbf{n}\bigtriangleup S - \vec{f}\bigtriangleup V \right\|_2$ $\leftarrow$ $\mathcal{I}_{mom}(G_e,G_c,\mathbf{C}, \bigtriangledown \mathbf{C},\mathbf{\theta}_{pde})$\;

\If{has pressure outlet}{$L_p=\left \|   [-p \mathbf{I}+ \mu \bigtriangledown \mathbf{u} ] \mathbf{n} +\hat{p_{0}} \mathbf{n}\right \|_2$ $\leftarrow$ $\mathcal{I}_{p} (G_e, G_c,\mathbf{N},\bigtriangledown \mathbf{N})$}
Optimizing parameters in $\mathcal{NN}$\;
\If{$L_{cont}<\epsilon \& L_{mom}<\epsilon$}{break}\Else{
      continue\;
    }
}
\end{algorithm}

\section{Results}\label{sec:numerical_validate}

In this section, we demonstrate the results obtained through inference after a \textbf{single} model is trained \textbf{once} under various conditions: two physical scenarios, multiple types of grids, multiple boundary conditions, and multiple source terms. The two types of scenarios are the Poisson equation and the steady incompressible flow problem (solved using the unsteady NS equations). Subsequent error calculations use the formula $RMAE= ({\textstyle \sum_{}^{}} \left | u^{predict}-u^{GT} \right | )/ ({\textstyle \sum_{}^{}} \left |u^{GT}  \right |)$, and for the steady incompressible flow problem, we also present velocity profiles along the streamlines and results for lift and drag coefficients, comparing with the unsupervised methods of the uniform MAC-grid\cite{wandel2020learning} and Spline-PINN\cite{wandel_spline-pinn_2022}.

In the process described above, we use the encoding of $\theta_{pde}$, along with the initial conditions and boundary conditions, to enable Gen-FVGN to differentiate inputs under different conditions. Therefore, this section primarily showcases the inference test results after training the Gen-FVGN under various meshes, different equations, different boundary conditions, and different source terms for PDEs solving. It is also important to emphasize that what follows can be divided into solving steady-state NS equation and Possion`s equation. In solving steady-state problems, we used only one model for a single training session. This means that the results for the subsequent Poisson equation and the steady-state NS equation are all derived from inference after training the same model.

\subsection{Possion's Equation}
The formula of the Poisson equation is shown in Eq.\eqref{eq:possion`s eq}. Here, $c$ represents a diffusion coefficient, and $f$ is the source term. When $f=0$, the Poisson equation can also be referred to as the Laplace equation (as shown in Tab.\ref{tab:possion-case}).

\begin{equation}\label{eq:possion`s eq}
    c\nabla^{2} \varphi=-f \quad in \quad \mathbf{\Omega}
\end{equation}

\begin{table}[!htbp]
\centering
\caption{Solving the Poisson equation and the wave equation involves setting the number of elements, boundary conditions, and the range of source terms. It's important to note that due to the substantial differences in the number of grid elements among various types, this table only represents an approximate average number of elements.}
\label{tab:possion-case}
\begin{tabular}{cccc}
\hline
\begin{tabular}[c]{@{}c@{}}Equation\\ Type\end{tabular} & Cells(\#avg.) & \begin{tabular}[c]{@{}c@{}}Dirichlet  BC\\ Range\end{tabular} & \begin{tabular}[c]{@{}c@{}}Source \\ Range\end{tabular} \\ \hline
Laplace                                                 & 8000          & \textless{}0,6\textgreater{}                             & -                                                       \\
Possion                                                 & 8000          & \textless{}0,6\textgreater{}                             & \textless{}1,4\textgreater{}   \\  \hline                       
\end{tabular}
\end{table}

\begin{figure}[!htbp]
\centering
\begin{minipage}{\linewidth}
    \includegraphics[width=1\textwidth]{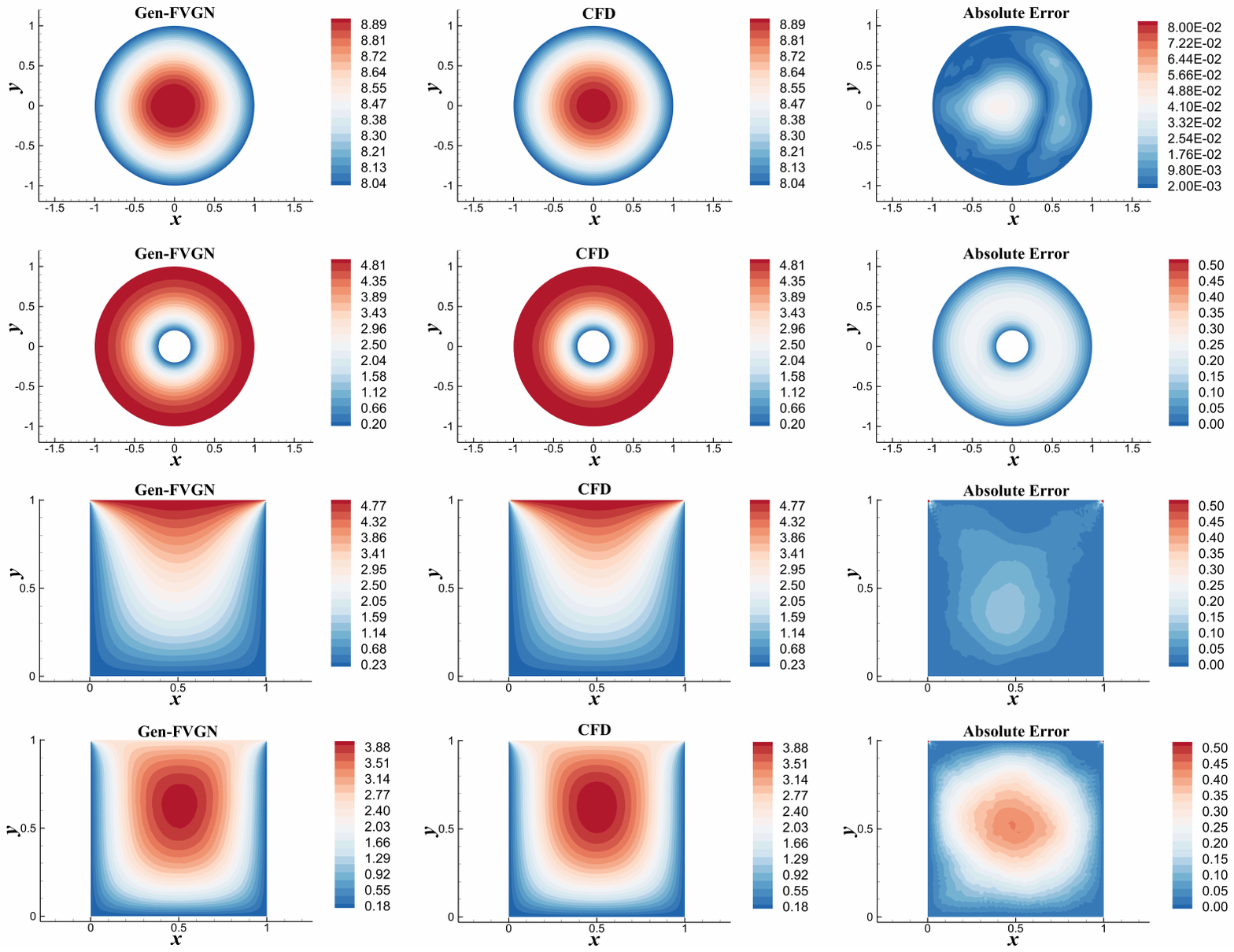}
    \end{minipage}
    \caption{Comparison of the solutions to the Poisson equation and the CFD results, where the first and second lines use the meshes from a) and b) in Fig.\ref{fig:circular_ring_mesh} respectively; the third and fourth lines use the meshes from a). and b). in Fig.\ref{fig:cavity_mesh} respectively.}
    \label{cavity_possion}
\end{figure}

We present the results of solving the Poisson equation on multiple different grids, where we employed completely different boundary conditions and source terms. In Fig.\ref{cavity_possion}, the first row of examples features a Dirichlet boundary condition applied at the outermost circle, with $\varphi=0 \quad in\quad \partial \Omega$; in the second row, both the inner and outer circles have Dirichlet boundary conditions with $\varphi=0 \quad in\quad \partial \Omega$ as well. The third and fourth rows feature a specified Type I boundary condition at the top, with a value greater than 0 assigned. The other three boundaries also have $\varphi=0 \quad in\quad \partial \Omega$. The results from Fig.\ref{cavity_possion} demonstrate that Gen-FVGN can achieve high accuracy in solving the Poisson equation.

As can be seen from Tab.\ref{Poisson-comp}, we have a significant advantage in solving capabilities compared to the MAC-grid method cited in \cite{wandel2020learning}. Our results were tested on the grid shown in Fig.\ref{fig:cavity_mesh}(e). The pixel-style grid of the MAC-grid method makes it difficult to solve for grids like those shown in Fig.\ref{fig:circular_ring_mesh}. Therefore, we set different source terms to compare the test results. Tab.\ref{Poisson-comp} indicates that the MAC-grid method, based on BC Loss, struggles to generalize with respect to the source terms, or perhaps cannot generalize at all. However, our method, which imposes boundaries rigidly, exhibits good generalizability. Gen-FVGN significantly outperforms the MAC-grid method across various metrics.

\begin{table}[!htbp]
\centering
\caption{Variant method solving poisson equation comparison, $RMAE= ({\textstyle \sum_{}^{}} \left | u^{predict}-u^{GT} \right | )/ ({\textstyle \sum_{}^{}} \left |u^{GT}  \right |)$ is Relative-Mean Absolute Error* for poisson flow field}
\label{Poisson-comp}
\begin{tabular}{c|cllclclclll}
\hline
\multirow{3}{*}{Method} &
  \multicolumn{3}{c}{Possion} &
  \multicolumn{2}{c}{Possion} &
  \multicolumn{2}{c}{Possion} &
  \multicolumn{4}{c}{Possion} \\
 &
  \multicolumn{3}{c}{c=0.1,source=0} &
  \multicolumn{2}{c}{c=0.1,source=1} &
  \multicolumn{2}{c}{c=0.1,source=2} &
  \multicolumn{4}{c}{c=0.1,source=3} \\
 & \multicolumn{11}{c}{RMAE $\times 10^{-3}$} \\ \hline
MAC-grid &
  \multicolumn{3}{c}{3.20} &
  \multicolumn{2}{c}{219.26} &
  \multicolumn{2}{c}{157.44} &
  \multicolumn{4}{c}{118.70} \\
\begin{tabular}[c]{@{}c@{}}Gen-FVGN\\ (ours)\end{tabular} &
  \multicolumn{3}{c}{1.41} &
  \multicolumn{2}{c}{2.55} &
  \multicolumn{2}{c}{2.78} &
  \multicolumn{4}{c}{4.88} \\ \hline
\end{tabular}
\\
\vspace{3pt} 
\caption*{\textbf{*}:Due to differences in the boundary condition application process between the one used in this paper and commercial solvers, the values of the upper-leftmost and upper-rightmost points were excluded when calculating the error for the cavity-mesh case in this paper.}
\end{table}

\subsection{Steady-State Navier-Strokes Equation}
In the previous section, this paper conducted an in-depth analysis and prediction of the various components of partial differential equations, including boundary conditions and source terms, achieving significant results. These results demonstrate that the adopted methods can be effectively applied to the solution processes of simple elliptic partial differential equations. However, the Navier-Stokes (NS) equations describing fluid mechanics behavior do not belong to either of the aforementioned types of equations and exhibit strong nonlinear characteristics. This subsection will employ the same neural network model as in the previous subsection to solve the incompressible NS equations, aiming to verify the capability and efficacy of the fully unsupervised neural network based on physical constraints in practical applications.

\begin{table}[!htbp]
\centering
\caption{Comparison of all NS case details used in this paper}
\label{tab:train-case}
\scalebox{0.97}{
\begin{tabular}{cccc|c}
\hline
\begin{tabular}[c]{@{}c@{}}Flow/Equation\\ Type\end{tabular} &
  Case &
  \begin{tabular}[c]{@{}c@{}}Average\\ Cell Count\end{tabular} &
  \begin{tabular}[c]{@{}c@{}}Reynolds Number Range\end{tabular} &
  \begin{tabular}[c]{@{}c@{}}Average Iteration\\ Steps During Training \end{tabular} \\ \hline
\multirow{3}{*}{Pipe Flow} &
  Cylinder &
  15000 &
  \multirow{3}{*}{{[}2,50{]}} &
  \multirow{10}{*}{500}  \\
                         & Square     & 8500  &                                               &  \\ \cline{1-4}
\multirow{3}{*}{Outflow} & NACA0012   & 12000 & \multirow{3}{*}{{[}1000,2000{]}}              &  \\
                         & RAE2822    & 12000 &                                               &  \\
                         & S809       & 12000 &                                               &  \\ \cline{1-4}
Cavity                   & Lid-driven & 8000  & {[}100,400{]}                                &  \\ \cline{1-4} \hline
\end{tabular}
}
\end{table}

This section showcases the various grids used in the subsequent solutions of the NS equations, as well as the range of boundary conditions, source terms, and iteration counts. In Tab.\ref{tab:train-case}, the velocity inlet profile for Pipe Flow uses a square distribution. The Outflow cases represent the external flow field examples, with uniform inflow. For instance, Fig.\ref{fig:farfield_mesh} shows the mesh for the external flow field, where the left boundary is the inlet boundary, and the top, bottom, and right boundaries are all pressure outlets. The grids used in this subsection also include not just those shown here, but also encompass the rectangular domain grids from the previous subsection, as shown in Fig.\ref{fig:cavity_mesh}.

Tab.\ref{tab:train-case} lists all considered cases and relevant details, including flow/equation type, average cell count, Reynolds number range, and average iteration steps during training. Fig.\ref{fig:cavity_mesh} to \ref{fig:pipeflow_mesh} display the mesh layouts for different cases, featuring a variety of grid types such as free quadrilateral, isotropic or anisotropic triangular, and quadrilateral meshes.

Solving the Navier-Stokes (NS) equations is a major challenge. Currently, the Physics-Informed Neural Networks (PINNs) approach struggles to solve equations under multiple boundary conditions in a single training process, or it can be said that the PINNs approach lacks generalizability for boundary conditions. While the methods based on MAC-grid\cite{wandel_teaching_2021,wandel2020learning} are relatively accurate and capable of solving with multiple boundary conditions, they exhibit significant errors in capturing the boundary layer flow fields. However, these issues are not present in the Generalized Finite Volume - Generalized Navier-Stokes (Gen-FVGN) method.

Due to the steady NS equations' high nonlinearity, solving the steady NS equations is more challenging than solving the Poisson equation. In Eq.\eqref{momtem}, the viscosity $\nu$ is set to 0.001 and the density $\rho$ is also set to 1, with the source term set to zero in the subsequent solutions of the NS equations. The variables are the grid and inlet boundary conditions. Of course, variations in $\nu$ and $\rho$ can also be included in $\mathbf{\theta_{pde}}$, allowing the model to learn the changes in various parameter combinations. However, this would expand the "dataset" range, thereby increasing the learning scope of the model. Due to considerations of computational memory overhead, changes in $\nu$ and $\rho$ were not considered in subsequent sections of this paper.

\subsubsection{Solution of Lid-Driven Cavity Flow}
Firstly, we conduct calculations for the lid-driven cavity flow, which is most commonly used for verifying the accuracy of traditional numerical algorithms. We have also compared the solution accuracy with the MAC grid method\cite{wandel2020learning,wandel2020teaching}. As shown in Tab.\ref{tab:Cavity-comp}, since the Navier-Stokes equations used in our tests no longer contain a source term, the MAC grid method based on BC Loss demonstrates higher solution accuracy. However, it still has slightly higher error in pressure solution compared to Gen-FVGN. This is mainly attributed to the use of MAC grid, where pressure and velocity are stored staggered, making the imposition of pressure boundary conditions challenging.
\begin{table}[!htbp]
\centering
\caption{Comparison of results by different methods in solving steady-state NS equations, $e_{U}= \frac{{\textstyle \sum_{}^{}} \left | u^{predict}-u^{GT} \right | + {\textstyle \sum_{}^{}} \left | v^{predict}-v^{GT} \right |}
{ {\textstyle \sum_{}^{}}\left |u^{GT}  \right | + {\textstyle \sum_{}^{}}\left |v^{GT}  \right |} $ is Relative Mean Absolute Error for velocity field, $e_{p}= \frac{{\textstyle \sum_{}^{}} \left | p^{predict}-p^{GT} \right |} { {\textstyle \sum_{}^{}}\left |p^{GT}  \right |} $ is Relative Mean Absolute Error for pressure field}
\label{tab:Cavity-comp}
\scalebox{1}{
\begin{tabular}{ccccccccc}
\hline
\multirow{3}{*}{\textbf{Method}} & \multicolumn{8}{c}{\textbf{Cavity Flow}}                                                                          \\ \cline{2-9} 
                                 & \multicolumn{2}{c}{Re=100} & \multicolumn{2}{c}{Re=200} & \multicolumn{2}{c}{Re=300} & \multicolumn{2}{c}{Re=400} \\
                                 & $e_{U}$      & $e_{p}$     & $e_{U}$      & $e_{p}$     & $e_{U}$      & $e_{p}$     & $e_{U}$      & $e_{p}$     \\ \hline
MAC-grid\cite{wandel2020learning}                         & 0.092        & 0.447       & 0.168        & 0.410       & 0.165        & 0.338       & 0.163        & 0.606       \\
Gen-FVGN                         & 0.034       & 0.021       & 0.0154      & 0.052       & 0.0192       & 0.0438      & 0.0263       & 0.065       \\ \hline
\end{tabular}
}
\end{table}

\begin{figure}[!htbp]
\centering
\begin{minipage}{\linewidth}\centering
    \includegraphics[width=1\textwidth]{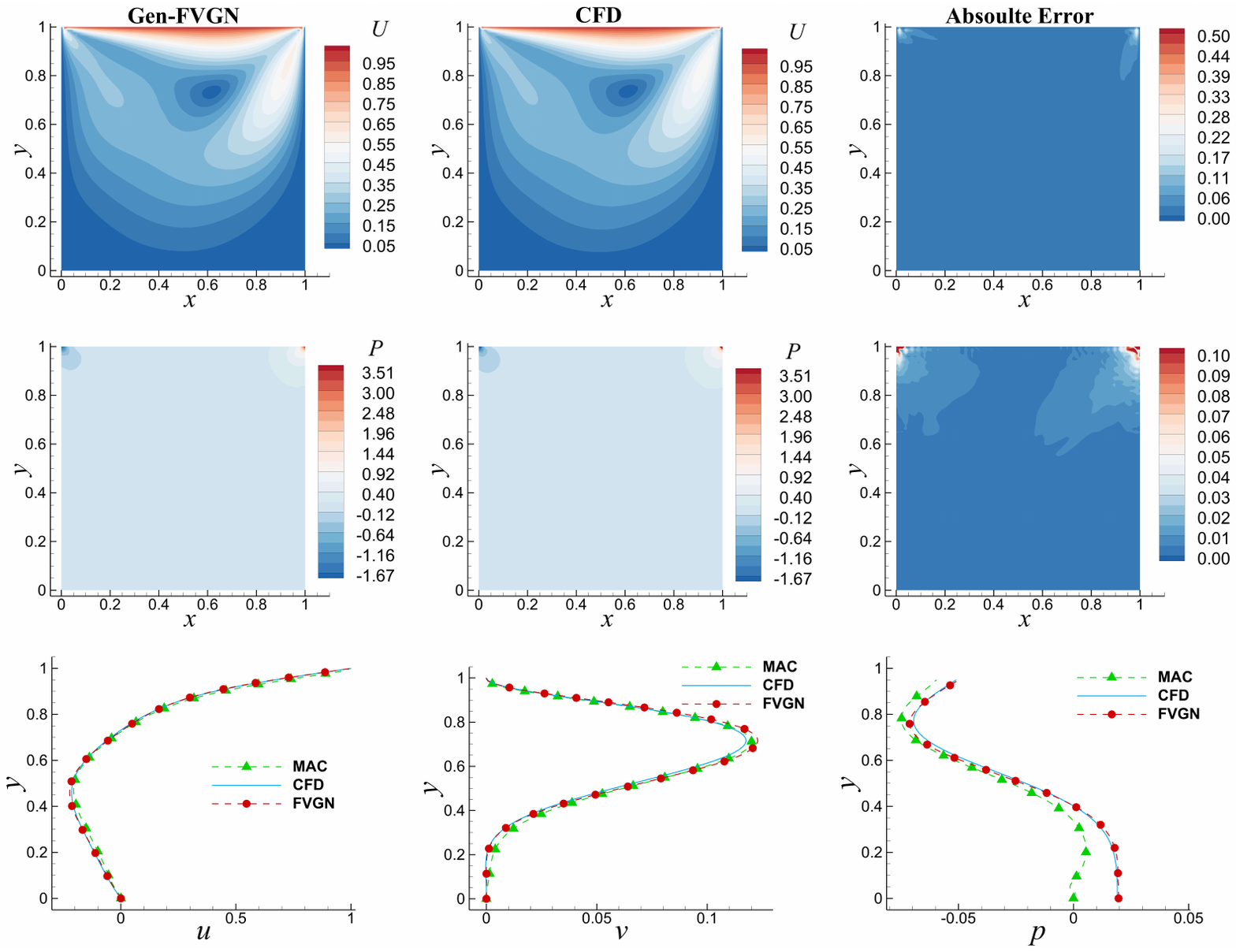}
    \end{minipage}
    \caption{Lid-driven flow field comparison, Re=100; Top: Comparison of combined velocity; Middle: Comparison of pressure fields; Bottom: Comparison of velocity and pressure field profiles along the vertical centerline of the square cavity}
    \label{cavity-flow-Re=100}
\end{figure}

\begin{figure}[!htbp]
\centering
\begin{minipage}{\linewidth}\centering
    \includegraphics[width=0.9\textwidth]{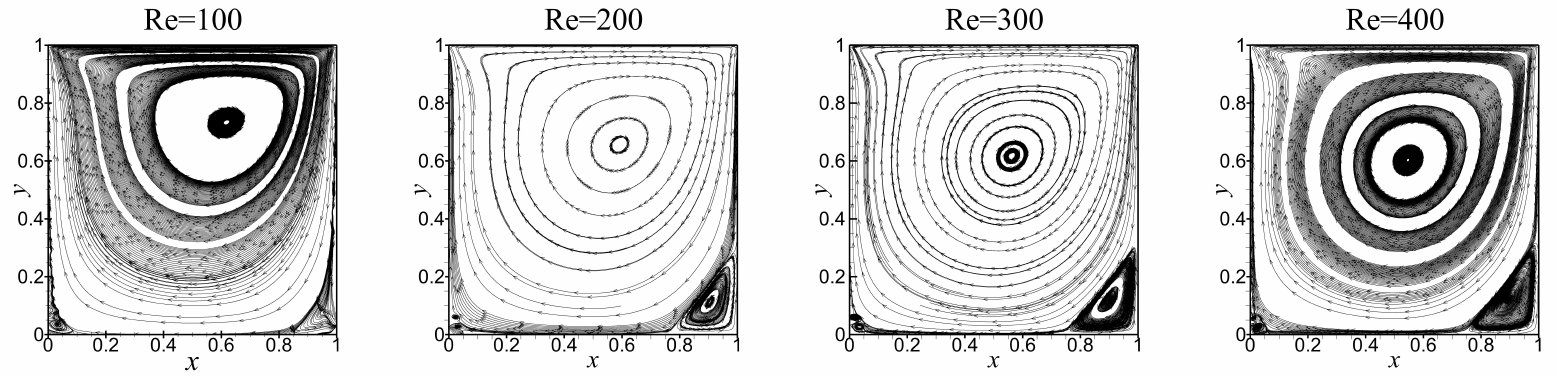}
    \end{minipage}
    \caption{Streamline diagrams of the solution results for lid-driven flow at different Reynolds numbers}
    \label{cavity-steamline-comp}
\end{figure}

From Tab.\ref{tab:Cavity-comp}, it is evident that Gen-FVGN demonstrates a remarkable performance in the accuracy of the velocity field across various Reynolds numbers, achieving precise predictions of the velocity field. However, the MAC grid method does not exhibit good generalization in the velocity field, performing relatively well at Re=100 but showing similar results at other Reynolds numbers. This underscores Gen-FVGN's superior generalization ability concerning boundary conditions.

We also present the flow field contour plots at Re=100, as shown in Fig.\ref{cavity-flow-Re=100}. Along the centerline, we have provided the profile distribution of the velocity and pressure fields. From the results in the last row, it can be observed that the MAC grid method still exhibits significant errors in pressure prediction. In contrast, Gen-FVGN maintains high solution accuracy, nearly overlapping with the CFD results.

Simultaneously, we provide the flow field streamline diagrams for Gen-FVGN at Re=100, 200, 300, and 400 in Fig.\ref{cavity-steamline-comp}. It is clear that Gen-FVGN possesses strong boundary generalization capabilities, accurately solving for the three vortex structures, particularly at Re=100, where it still identifies three vortex structures. This further highlights the boundary generalization capabilities of the Gen-FVGN method. Indeed, the lid-driven cavity flow example is one of the most classic test cases for assessing the accuracy of computational algorithms in the CFD field. The accuracy of solutions in this test largely predicts the algorithm's performance on other test cases, especially in resolving small-scale vortex structures. From the results discussed, it is evident that Gen-FVGN meets the solution accuracy requirements of second-order finite volume methods. Indirectly, this also shows that the GN blocks used as the backbone network in this method have good nonlinear fitting capabilities. Additionally, its Encoder-Processor-Decoder architecture provides the model with the ability to distinguish between different Reynolds numbers or boundary conditions effectively.

\subsubsection{Steady Flow Around a Cylinder}
\begin{figure}[!htbp]
\centering
\begin{minipage}{\linewidth}\centering
    \includegraphics[width=1\textwidth]{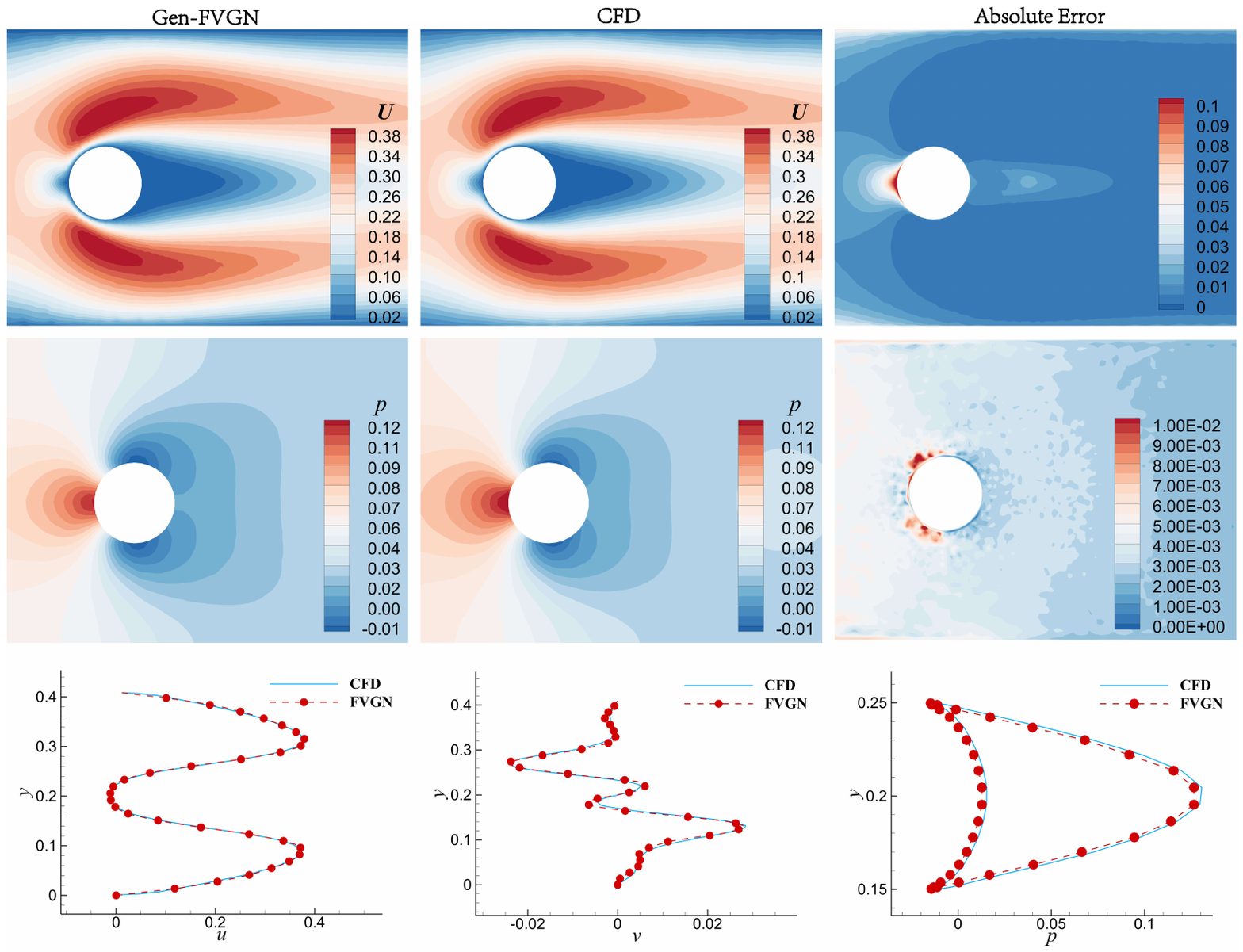}
    \end{minipage}
    \caption{Gen-FVGN steady-state flow solution around a cylinder; Top: Comparison of combined velocity; Middle: Comparison of pressure fields; Bottom: Comparison of velocity and pressure field profiles along a tangent line in the vertical direction behind the cylinder; Bottom right: Pressure distribution curve on the surface of the cylinder}
    \label{fig:Gen-FVGN-cylinder-steady-poly}
\end{figure}

Introducing obstacles in the domain increases the complexity of the solution. Here, we present the solution results for the steady flow around a cylinder using Gen-FVGN in the computational domain. Notably, these results are obtained on a polygonal mesh, as shown in Fig.\ref{fig:pipeflow_mesh}(d). In such cases, we pay particular attention to the accuracy near the boundary layer, especially the pressure distribution on the surface of the cylinder. From the bottom right of Fig.\ref{fig:Gen-FVGN-cylinder-steady-poly}, it can be seen that Gen-FVGN achieves very high accuracy in solving pressure distribution. Additionally, we have cut a vertical cross-section 0.1m behind the cylinder and provided the velocity distribution along this cross-section, as shown in the first two images of the third row of Fig.\ref{fig:Gen-FVGN-cylinder-steady-poly}.

\begin{table}[!htbp]
\centering
\caption{Variant method solving steady state cylinder flow comparsion}
\label{tab:Cylinder-comp}
\begin{tabular}{cccl|ccl|ccl}
\hline
\multirow{3}{*}{Method} & \multicolumn{3}{c|}{Cylinder Flow}          & \multicolumn{3}{c|}{Cylinder Flow}            & \multicolumn{3}{c}{Cylinder Flow}           \\
              & \multicolumn{3}{c|}{Re=2}                & \multicolumn{3}{c|}{Re=20}                 & \multicolumn{3}{c}{Re=40}              \\
              & $C_d$      & \multicolumn{2}{c|}{$C_l$}  & $C_d$      & \multicolumn{2}{c|}{$C_l$}    & $C_d$      & \multicolumn{2}{c}{$C_l$} \\ \hline
Spline-PINN\cite{wandel_spline-pinn_2022}   & 29.7       & \multicolumn{2}{c|}{-0.456} & 4.7        & \multicolumn{2}{c|}{5.64e-04} & -          & \multicolumn{2}{c}{-}     \\
Gen-FVGN(ours)          & 31.04          & \multicolumn{2}{c|}{0.677} & 5.323          & \multicolumn{2}{c|}{0.0104}  & 4.4405        & \multicolumn{2}{c}{0.01345} \\
CFD                     & 32.035         & \multicolumn{2}{c|}{0.774} & 5.57020        & \multicolumn{2}{c|}{0.00786} & 4.8405        & \multicolumn{2}{c}{0.00223} \\
DFG-Benchmark\cite{DFG_bench} & -          & \multicolumn{2}{c|}{-}      & 5.58       & \multicolumn{2}{c|}{0.0106}   & -          & \multicolumn{2}{c}{-}  \\ \hline  
\end{tabular}
\end{table}

From Fig.\ref{fig:Gen-FVGN-cylinder-steady-poly}, it can also be observed that Gen-FVGN has an advantage in the accuracy of velocity solutions. Moreover, this example is solved on a polygonal mesh with a Reynolds number of 20. The velocity field results of Gen-FVGN are particularly close to the CFD results, as can be seen from the first two images in the third row, where the profile prediction of the velocity field nearly coincides with the CFD results.

At the same time, the accurate solution of the pressure distribution directly determines the accuracy of the lift and drag coefficients. The accuracy of the aerodynamic coefficients directly determines the usability of this method. From Tab.\ref{tab:Cylinder-comp}, it can be seen that under steady conditions, the aerodynamic force prediction results of Gen-FVGN are quite precise. We used the DFG benchmark \cite{DFG_bench} as a standard of accuracy, which is obtained using a high-order finite element method. Our method is theoretically second-order, similar to the finite difference method used with MAC grids. However, due to the characteristics of the polygonal mesh, it has fewer cells and more vertices. Therefore, we obtained better results with a sparser mesh. We also compared our results with those of commercial software (as shown in the CFD row of Tab.\ref{tab:Cylinder-comp}). Our method can match commercial software in predicting lift coefficients, but there is still a gap in drag coefficients. We believe this is mainly due to the convergence characteristics of the neural network. We simultaneously solved multiple grids, equations, and boundary conditions, which pose higher coding demands on our network. It requires accurate differentiation of flow field distributions under different conditions. Therefore, the main reason for the drop in the accuracy of the drag coefficient is due to the insufficient accuracy of the pressure field prediction. In subsequent work, we will focus on improving the predictive capability of Gen-FVGN in this area.

\subsubsection{Low-Speed Airfoil Flow}\label{sec:low-speed-airfoil}

In facing more complex wing flow problems, effectively advancing time presents a significant challenge. Gen-FVGN employs an Implicit-Explicit (IMEX) time scheme. However, to ensure that each time step progresses to a convergent outcome, additional inner iterations are required. This essentially involves adding an extra judgment(details can be seen in Algorithm~\ref{alg:train_process}) within the time step cycle to verify whether the residual of the equations, predicted by the current neural network for the physical fields of the next time step, has decreased to a specified magnitude. If so, the results of this iteration are returned to the training pool for further time progression; if not, they are not returned to the training pool.
\begin{figure}[!htbp]
\centering
\begin{minipage}{\linewidth}\centering
    \includegraphics[width=1\textwidth]{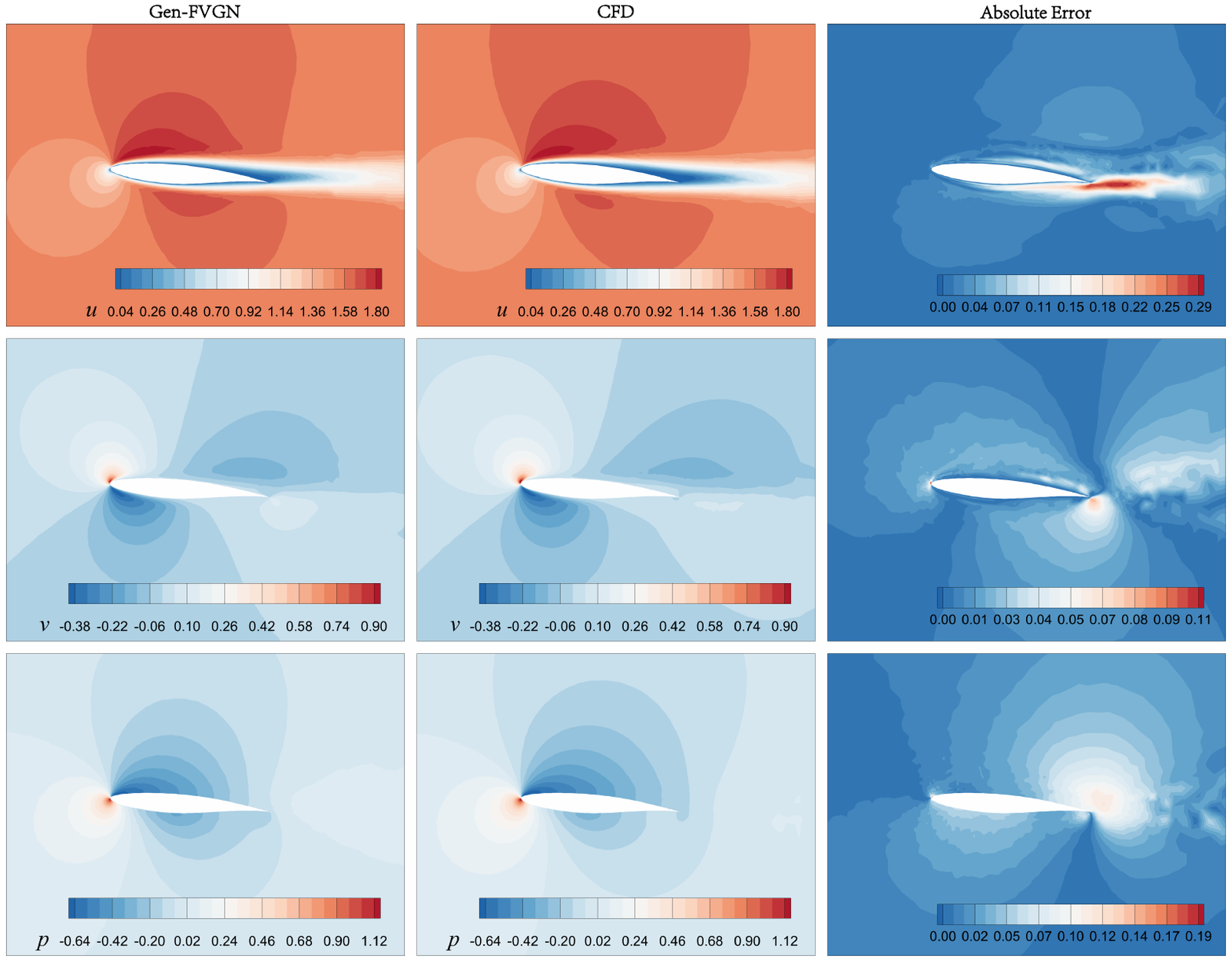}
    \end{minipage}
    \caption{Gen-FVGN solution for the steady-state flow field around RAE2822, Re=1550}
    \label{fig:Gen-FVGN-Rae282-steady}
\end{figure}

Gen-FVGN not only achieves good prediction results in pipe flow, but also demonstrates commendable performance in external flow fields. However, it must be acknowledged that there is a noticeable difference between the pressure field predictions by Gen-FVGN and the CFD simulation results, as seen in Fig.\ref{fig:Gen-FVGN-Rae282-steady}. The test case in question involves the RAE2822 airfoil at an angle of attack of 2.3° with a Reynolds number of 1550. It is also important to note that the Reynolds number in this case is markedly different from that of the lid-driven cavity flow previously discussed. Therefore, this situation also tests the capability of Gen-FVGN to handle completely different solution domains, and both types of cases are solved and tested through inference from the same model after a single training session.

\begin{figure}[!htbp]
\centering
\begin{minipage}{\linewidth}\centering
    \includegraphics[width=1\textwidth]{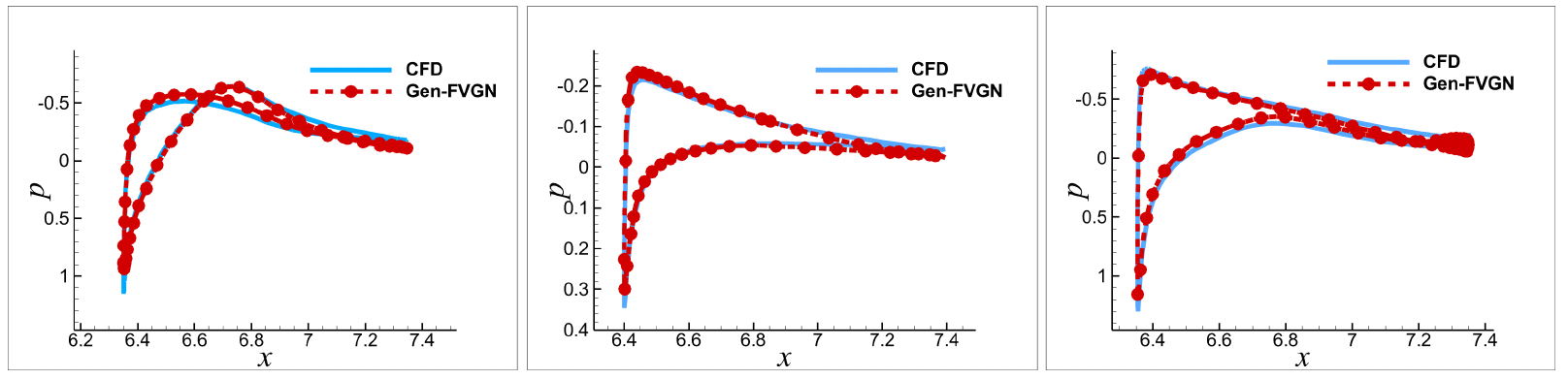}
    \end{minipage}
    \caption{Gen-FVGN to solve the pressure distribution curves on the surfaces of various airfoils; Left: S809 airfoil, Re=1440, AoA=3°; Middle: NACA0012 airfoil, Re=770, AoA=3°; Right: RAE2822 airfoil, Re=1550, AoA=3°}
    \label{fig:Gen-FVGN-Rae282-steady-p-dist}
\end{figure}

The final velocity field prediction RMAE for this example is $0.032$, and the pressure field prediction RMAE is $0.0784$. This demonstrates a consistent result with other scenarios, such as the lid-driven flow or flow around a cylinder, where the prediction accuracy of the pressure field is always lower than that of the velocity field. This is primarily due to the particular characteristics of the incompressible Navier-Stokes equations. Therefore, enhancing the model's representational capacity will be our main focus in subsequent work. We believe that the cause of this issue may be due to a mismatch between the size of the neural network's hidden layer space and the number of actual problems being solved, or perhaps, we have placed too much "learning pressure" on the neural network. From the content in Tab.\ref{tab:train-case}, we can estimate that Gen-FVGN actually faces up to tens of thousands of steady problems. Gen-FVGN maintains the same hidden space size as FVGN, which uses 15 message-passing layers and a hidden layer size of 128. Gen-FVGN uses 12 message-passing layers with a hidden layer size of 128. We reduced the size of the message-passing layers to balance the ratio of GPU memory and batch size because too small a batch size would prevent Gen-FVGN from achieving optimal solution accuracy.

\subsection{Efficency Comparision}
Finally, similar to the validation method for FVGN, we tested the efficiency improvement of Gen-FVGN compared to traditional algorithms. Here, we did not directly provide specific time improvement ratios but instead presented a comparison of the number of iterations required to solve the equations. We believe that since Gen-FVGN is an entirely GPU-based algorithm and traditional commercial software is generally still based on MPI-based CPU parallel architectures, directly comparing the two types of methods could inevitably introduce other errors. Therefore, we consider directly comparing the number of equation iterations to be a more balanced approach.
\begin{figure*}[!htbp]
\centering
\begin{minipage}{\linewidth}\centering
    \includegraphics[width=1\textwidth]{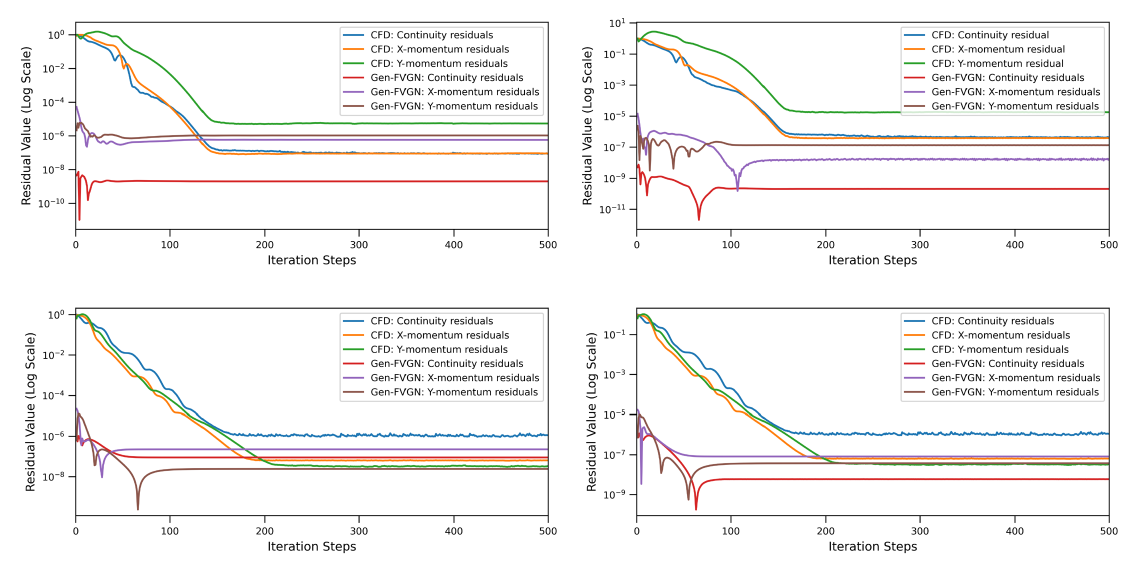}
    \end{minipage}
    \caption{Resdiual Comparsion; left up: Cavtiy lid-driven case Re=100; right up Cavity lid-driven case Re=400; left down: Cylinder flow case Re=20; right down Cylinder flow case Re=40}
    \label{fig:Gen-FVGN-efficency}
\end{figure*}
In this context, we present the residual variation graphs after convergence to steady-state results for Gen-FVGN when solving the steady lid-driven cavity flow and cylinder flow around problems. From Fig.\ref{fig:Gen-FVGN-efficency}, it can be seen that during the inference process, Gen-FVGN maintains the stability of steady problems very well, and almost no residual changes subsequently. Most importantly, compared to commercial software, Gen-FVGN requires only about half the number of iterations to converge in almost all test cases. Considering the natural advantages of GPU architecture, the final speed improvement achieved by Gen-FVGN in solving is expected to be quite remarkable. Even more impressively, as shown in the bottom left of Fig.\ref{fig:Gen-FVGN-efficency}, Gen-FVGN can reach a converged state with only 1/4 the number of iterations required by commercial software. Moreover, the final convergence residuals of all results are comparable to those achieved with commercial software.

\section{Discussion and Conclusion}
This paper has implemented a model that uses only one model to train and solve different PDEs, supplemented with various boundary conditions and source terms. The combination of the finite volume method and graph neural networks also allows the model proposed in this paper to adapt to various types of two-dimensional unstructured grids. However, it must be pointed out that this paper only conducted solution verification on the Poisson equation and the steady NS equation, and it has not truly extended to the unsteady NS equation with vortex shedding phenomena. This is mainly because solving unsteady NS equations under multiple boundary conditions simultaneously is very challenging. Existing PINN-like methods already have strong capabilities in solving under a single boundary condition, and our method still faces some difficulties in training and solving unsteady problems under multiple boundary conditions in an unsupervised manner. Nevertheless, we believe that the method proposed in this paper provides an excellent platform, seamlessly integrating traditional numerical methods with deep learning models, thereby better promoting the integration and development of adjacent disciplines.

To address the issue of high dependency on high-precision datasets, this paper successfully developed an improved unsupervised training algorithm based on the deep learning framework and utilizing the advantages of GPU parallel computing. This algorithm implements a complete, differentiable finite volume method, including efficient integral and gradient reconstruction algorithms, allowing the model to directly solve physical equations during training without relying on pre-computed data. The core of this method is its ability to handle multiple boundary conditions in a single training iteration, significantly enhancing the model's generalization capabilities and application scope. Through this unsupervised learning approach, our model not only demonstrated excellent mesh generalization but also showed significant improvements in generalizing boundary conditions and PDEs. This progress provides a new approach to solving complex physical equations using graph neural networks, promising widespread application in computational fluid dynamics and broader engineering fields. Moreover, compared to the MAC method based on finite difference algorithms, our model exhibited better generalization capabilities and solution accuracy comparable to commercial CFD software.

\section*{Code and data availability}
The code used in this article can be found in the GitHub repository: \url{https://github.com/Litianyu141/Gen-FVGN-steady}.

\section*{Acknowledgements}
This research is funded by the National Key Project of China (Grant No. GJXM92579) and is also supported by the Sichuan Science and Technology Program (Project No. 2023YFG0158).
 \clearpage
 \bibliographystyle{elsarticle-num} 
 \bibliography{cas-refs}
 
 \clearpage

\appendix

\section{Gradient Reconstruction}\label{sec:grad-verify}
Following the description of the loss function construction process, it is evident that gradient reconstruction is the most crucial process in the entire discretization discussed in this paper. As described in Ref.\cite{moukalled2016finite}, gradient reconstruction on unstructured grids can generally be divided into two methods: the Green-Gauss method (abbreviated as GG) and the Least Squares method (LSQ). The GG method can be considered an extension of the divergence theorem, and as mentioned earlier, the divergence theorem on unstructured grids poses certain difficulties for neural network learning. Therefore, here we adopt the Weighted Least Squares method (WLSQ), which is more widely used in practical engineering. This method is derived from Taylor's formula, and the specific process can be referred to in Ref.\cite{moukalled2016finite}. Here, we directly present the specific formula for the Weighted Least Squares method in two dimensions, i.e., the matrix form of the normal equations as given in Eq.\eqref{lsq}:

\begin{equation}\label{eq:wlsq}
\left[\begin{array}{cc}
\sum_{i} w_{i}^{2} \Delta x_{i}^{2} & \sum_{i} w_{i}^{2} \Delta x_{i} \Delta y_{i} \\
\sum_{i} w_{i}^{2} \Delta x_{i} \Delta y_{i} & \sum_{i} w_{i}^{2} \Delta y_{i}^{2}
\end{array}\right]\left[\begin{array}{c}
\frac{\partial \phi}{\partial x}  \\
\frac{\partial \phi}{\partial y} 
\end{array}\right]=\left[\begin{array}{c}
\sum_{i} w_{i}^{2} \Delta x_{i}\left(\phi_{i}-\phi_{0}\right) \\
\sum_{i} w_{i}^{2} \Delta y_{i}\left(\phi_{i}-\phi_{0}\right)
\end{array}\right]
\end{equation}

In Eq.\eqref{eq:wlsq}, $w$ represents the inverse distance weight, and $\left [\frac{\partial \phi}{\partial x} ,\frac{\partial \phi}{\partial y} \right ]^T$ is the gradient at the point $(x_0,y_0)$. $\phi_{i} $ represents the physical quantity at other vertices adjacent to the current vertex $(x_0,y_0)$, or it can be referred to as the stencil for constructing the gradient at $(x_0,y_0)$.

\begin{figure}[!htbp]
\begin{minipage}{\linewidth}
    \centering
    \includegraphics[width=0.7\textwidth]{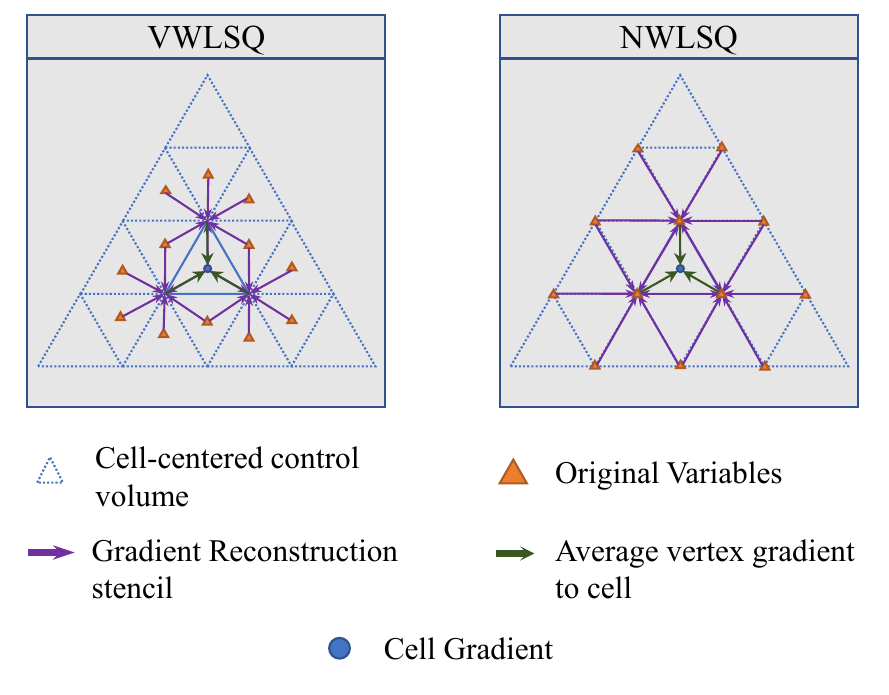}
    \caption{Comparison of templates for two gradient reconstruction methods, left: VWLSQ\cite{zhang2017vertex,chen2022vertex}; right: the gradient reconstruction template used in this paper.}
    \label{fig:grad-rec-stencil}
\end{minipage}
\end{figure}
From the analyses by Ref.\cite{diskin2010comparison,wang2019accuracy}, it is known that the WLSQ method guarantees first-order accuracy, meaning that it achieves at least first-order accuracy on any mesh. Therefore, after reconstructing $\bigtriangledown \mathbf{u}_v$ using the WLSQ, we use a simple arithmetic mean method to average the gradients of physical quantities at the vertices forming a cell to the cell center, $\mathbf{u}_{c_j} = \frac{1}{N}\sum_{v_i \in cell_j} \bigtriangledown \mathbf{u}_{v_i}$. Adding the two gradient components then yields the result of the divergence operator. This method can be seen as an approximation of the gradient on unstructured grids using finite differences, as the right-hand side of Eq.\eqref{eq:wlsq} is essentially the difference between the physical quantities at the current vertex and the surrounding vertices. Meanwhile, averaging the vertex gradients to the cell center is also mentioned in the works by Ref.\cite{zhang2017vertex,chen2022vertex} (VWLSQ). However, it must be noted that VWLSQ still places the physical variables at the cell center, albeit constructing the final least squares equations at the vertices (as shown in Fig.\ref{fig:grad-rec-stencil}). Therefore, there are some differences in the gradient reconstruction accuracy validation between the original variables given in VWLSQ and the method used in this paper, although both ultimately use arithmetic averaging to interpolate the vertex gradients back to the cell center. In light of this, we have still conducted a verification of the reconstruction accuracy of the algorithm used in this paper. Sec.\ref{sec:grad-verify} provides a detailed description of this process. Finally, it should be noted that due to space limitations in the article, we have not presented the process of solving the conservative form NS equations using the divergence theorem discretization, not because it cannot converge, but due to its low efficiency. Therefore, subsequent numerical validation examples only demonstrate the NS equation solution results for the convective terms using the finite difference-based WLSQ method.

\begin{figure}[!htbp]
\begin{minipage}{\linewidth}
    \centering
    \includegraphics[width=0.7\textwidth]{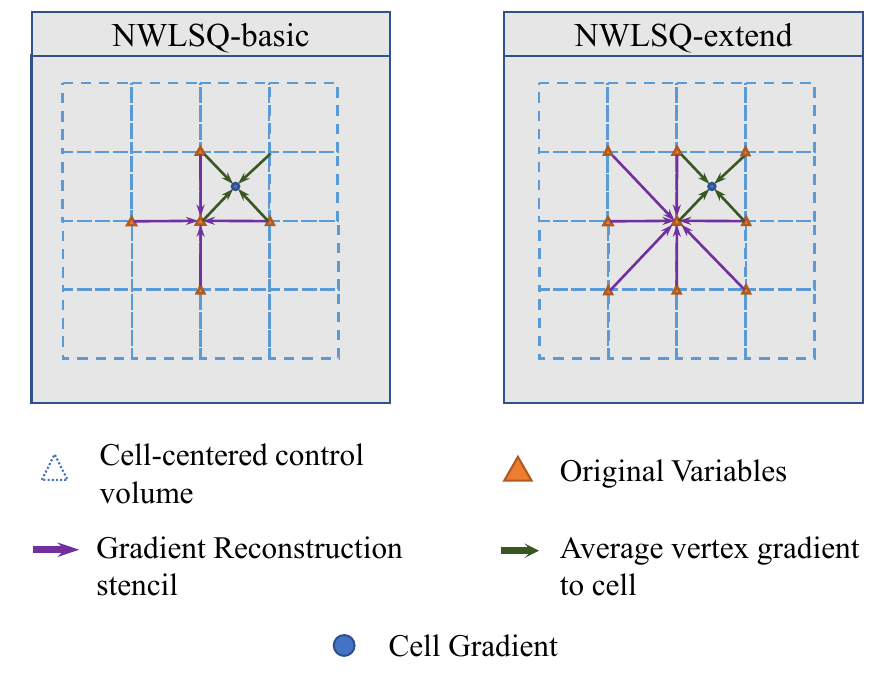}
    \caption{The two gradient reconstruction templates used in this paper are, on the left: NWLSQ-basic, which means the basic template; on the right: NWLSQ-extend, which means the extended template.}
    \label{fig:grad-rec-stencil-basic-ext}
\end{minipage}
\end{figure}

In this paper, the gradient reconstruction stencils are categorized into basic and extended templates. The basic template involves calculating WLSQ using only the first-order neighbors of each vertex. The extended template, on the other hand, not only selects the first-order neighbors of the vertex but also includes all the vertices of the cells sharing the same vertex, as shown on the right side of Fig.\ref{fig:grad-rec-stencil-basic-ext}. In triangular meshes, the basic and extended templates are equivalent. However, in quadrilateral and polygonal meshes, the size of the extended template is significantly larger than that of the basic template. Especially in polygonal meshes, each vertex's basic template only has three neighboring vertices, making it equivalent to a cell-centered triangular mesh least squares gradient reconstruction template. The gradient at the vertices can then be interpolated to the cell center to construct the various terms in the previously mentioned NS equations.

For particularly orthogonal quadrilateral meshes, such as the mesh shown in Fig.\ref{fig:grad-rec-stencil-basic-ext} or Fig.\ref{fig:grad_rec_mesh}a, both LSQ and WLSQ can achieve second-order accuracy because, in these cases, the least squares method can be considered a special case of the central difference format. However, from the analysis of Ref.\cite{diskin2010comparison,wang2019accuracy}, it is known that the LSQ, GG, or WLSQ methods only achieve first-order accuracy on triangular meshes with the classical variable arrangement.

Yet, our method can achieve second-order accuracy at the cell level on any mesh. To verify the theoretical analysis and conclusions of the gradient reconstruction method and stencil analysis, we conducted numerical accuracy verification on four different types of typical isotropic and anisotropic meshes. Specifically, since this section only validates the feasibility of the gradient reconstruction template, the accuracy comparison was performed only on high-quality meshes without a detailed comparison on disturbed anisotropic meshes. However, the numerical examples in Sec.\ref{sec:numerical_validate} also demonstrate that our method can achieve good results on any type of two-dimensional mesh.

\subsection{Gradient Reconstruction Accuracy Test}
\begin{figure}[!htbp]
\begin{minipage}{\linewidth}
    \centering
    \includegraphics[width=1\textwidth]{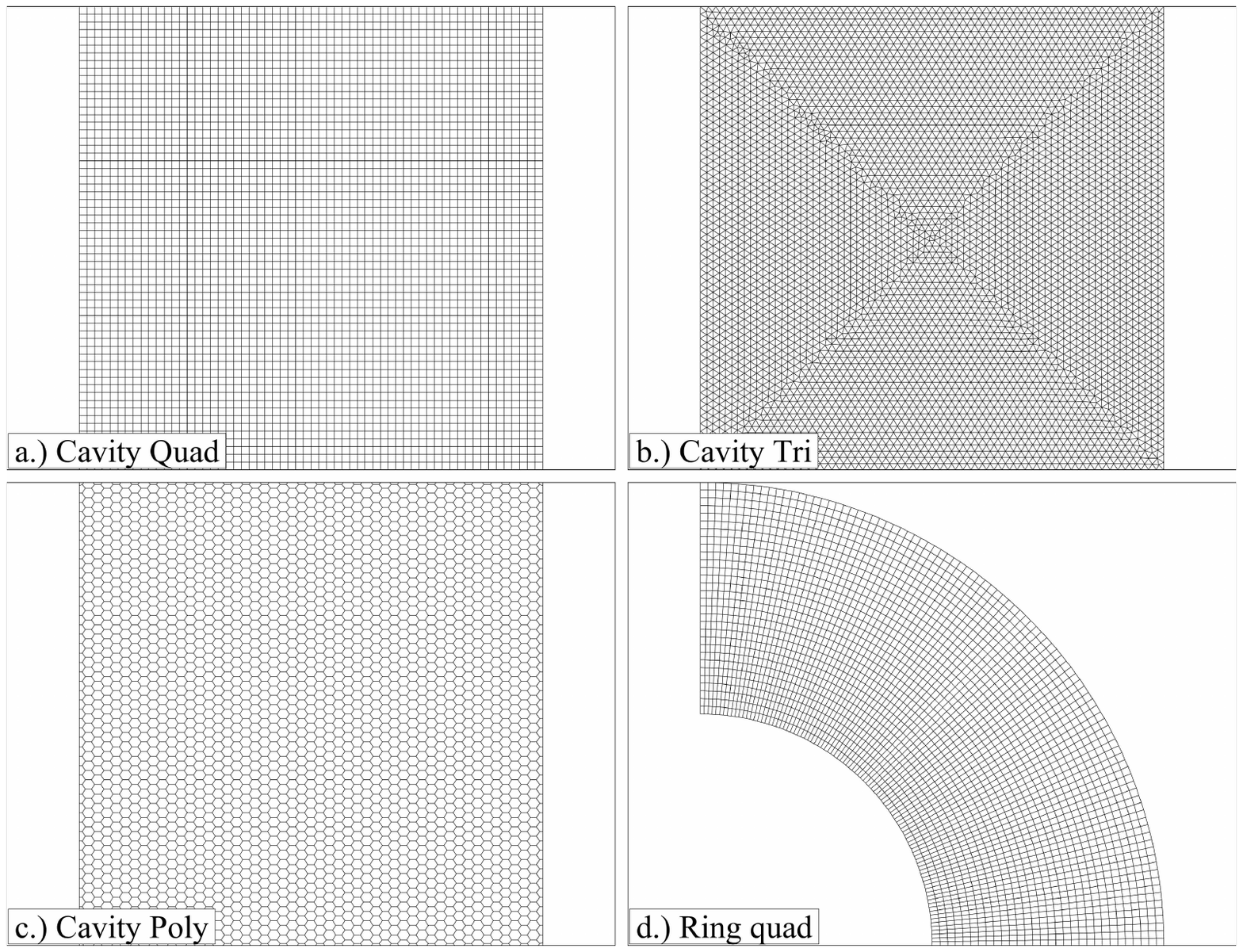}
    \caption{The gradient reconstruction accuracy tests used four different types of meshes: a) isotropic square quadrilateral, b) isotropic square triangle, c) isotropic square polygon, d) isotropic annular quadrilateral.}
    \label{fig:grad_rec_mesh}
\end{minipage}
\end{figure}

The form of the test function used for gradient reconstruction on isotropic grids is:
\begin{equation}\label{eq:grad_eular_eq}
    f(x, y)=x+y+C_{x} \sin \left(\dfrac{\pi x}{L_{x r e f}}\right)+C_{y} \sin \left(\dfrac{\pi y}{L_{y r e f}}\right)+C_{x y} \cos \left(\dfrac{\pi x y}{L_{x r e f} L_{y r e f}}\right)
\end{equation}
The flow field function uses an infinitely smooth nonlinear function similar to the scalar Euler manufacturing solution, and introduces reference scales $L_{xref}$ and $L_{yref}$, weighting coefficients $C_x, C_y, C_{xy}$, and linear terms based on the scalar Euler manufacturing solution function to ensure that the gradient field does not have zero points. It can simulate various smooth gradient fields and can select reference scales and weighting coefficients according to different testing requirements. In this section, we set $C_x=C_y=C_{xy}=1.0$, $L_{xref}=L_{yref}=15$. We use traditional mesh refinement methods and mesh reduction methods to test the accuracy and order of accuracy of the GG-Cell gradient reconstruction method at the vertices and cell centers of the computational mesh shown in Fig.\ref{fig:grad_rec_mesh}.

In fact, this paper has verified various gradient reconstruction algorithms as well as the accuracy of reconstructing variable gradients at different positions. Since the variables were decoded to the grid vertices in this paper, the original variables can be considered to be located at the Grid vertices. However, the gradients needed are at the cell centers, thus leading to the various reconstruction algorithms shown in Tab.\ref{tab:grad_rec_method}.
This table lists and characterizes various gradient reconstruction algorithms used in this paper:

\begin{sidewaystable}
\centering
\caption{The various gradient reconstruction algorithms used in this paper are briefly named and characterized as follows:}
\label{tab:grad_rec_method}
\renewcommand{\arraystretch}{1.8} %
\scalebox{0.8}{
\begin{tabular}{ccccc}
\hline
\textbf{\begin{tabular}[c]{@{}c@{}}Reconstruction\\ Algorithm\end{tabular}}    & \textbf{\begin{tabular}[c]{@{}c@{}}Original Variable\\ Position\end{tabular}} & \textbf{\begin{tabular}[c]{@{}c@{}}Gradient Reconstruction\\ Position\end{tabular}} & \textbf{\begin{tabular}[c]{@{}c@{}}Average\end{tabular}} & \textbf{Features}                                                                                 \\ \hline
GG-N                                                        & Grid vertices                                                         & Cell centers                                                         & No                                                        & Unweighted, basic template                                                                                    \\
WSLQ-N                                                      & Grid vertices                                                         & Grid vertices                                                         & No                                                        & Inverse distance weighting, basic template                                                                                 \\
WSLQ-NX                                                     & Grid vertices                                                         & Grid vertices                                                         & No                                                        & Inverse distance weighting, extended template                                                                                 \\
WSLQ-NX-N                                                   & Grid vertices                                                         & Grid vertices                                                         & No                                                        & \begin{tabular}[c]{@{}c@{}}1st: Inverse distance weighting, extended template\\ 2nd: Inverse distance weighting, basic template\end{tabular}                   \\
WSLQ-NX-NX                                                  & Grid vertices                                                         & Grid vertices                                                         & No                                                        & \begin{tabular}[c]{@{}c@{}}1st: Inverse distance weighting, extended template\\ 2nd: Inverse distance weighting, extended template\end{tabular}                   \\ \hline
\begin{tabular}[c]{@{}c@{}}WLSQ-NX-\\ CX\end{tabular}       & Grid vertices                                                         & Cell centers                                                         & No                                                        & \begin{tabular}[c]{@{}c@{}}1st: Inverse distance weighting, extended template\\ 2nd: Inverse distance weighting, reconstruct second-order derivatives interpolation to cell centers\end{tabular}            \\
\begin{tabular}[c]{@{}c@{}}WLSQ-NX-\\ avgC\end{tabular}     & Grid vertices                                                         & Cell centers                                                         & Yes                                                        & \begin{tabular}[c]{@{}c@{}}Inverse distance weighting, extended template,\\ arithmetic average to cell centers\end{tabular}                              \\
\begin{tabular}[c]{@{}c@{}}WLSQ-NX-\\ NCX-avgC\end{tabular} & Grid vertices                                                         & Cell centers                                                         & Yes                                                        & \begin{tabular}[c]{@{}c@{}}1st: Inverse distance weighting, extended template\\ 2nd: Inverse distance weighting, extended template with average values at cell centers\\ Arithmetic average to cell centers\end{tabular} \\
\begin{tabular}[c]{@{}c@{}}WLSQ-NX-\\ NC-avgC\end{tabular}  & Grid vertices                                                         & Cell centers                                                         & Yes                                                        & \begin{tabular}[c]{@{}c@{}}1st: Inverse distance weighting, extended template\\ 2nd: Inverse distance weighting, basic template with average values at cell centers\\ Arithmetic average to cell centers\end{tabular} \\ \hline
\end{tabular}
}
\end{sidewaystable}

From the Tab.\ref{tab:grad_rec_method}, it is evident that the paper also adopted a method of reconstructing the gradient twice to reduce the reconstruction error. Here, 'N' refers to the basic template, similar to the one shown on the right side of Fig.\ref{fig:grad-rec-stencil}, where the gradient template for each grid point includes all the vertices adjacent at first order. 'NX' specifically refers to the extended template, especially for quadrilateral or polygonal meshes, where the template for each grid point includes not only its first-order neighbors but also all the vertices of the cells that share a vertex with the current point. In the most extreme case, this might include third-order neighbors in polygonal meshes.

The aforementioned second reconstruction refers to the process where, after the initial reconstruction of the gradient at the grid points, the original variable is interpolated to the cell center using the following formula:

\begin{equation}\label{eq:inteplot-2nd}
    \phi_c=\dfrac{1}{N} \sum_{v \in cell_i}^{} (\phi_v+\overrightarrow{r} \cdot\bigtriangledown \phi_v)
\end{equation}
During the second reconstruction, variables at the cell centers that share a point with the grid point are introduced as part of the template. The last method, which is used in subsequent numerical example validations in this paper, is denoted by WLSQ-NX-CX. It involves first using an extended template to reconstruct the gradient $\nabla \phi_v$ at the grid points. Then, the gradient of the gradient $\nabla(\nabla \phi_v)$ is reconstructed using the extended template WSLQ. The first-order gradient is treated as the original variable and interpolated to the cell center using the formula \eqref{eq:inteplot-2nd}. After this, the gradient error based at the cell center is computed.

\begin{figure}[!htbp]
	\centering
	\begin{subfigure}{0.49\linewidth}
		\centering
		\includegraphics[width=\linewidth]{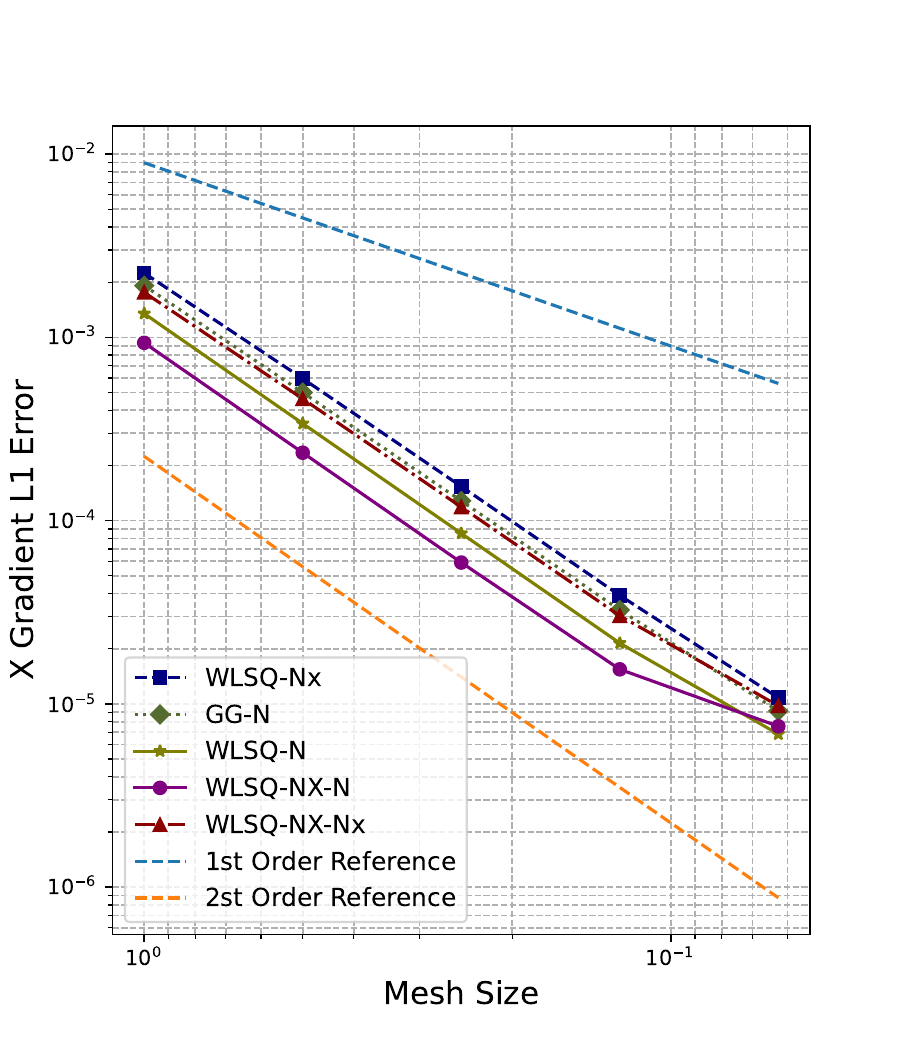}
		\caption{Accuracy of gradient reconstruction in the x-direction for a square cavity quadrilateral.}
	\end{subfigure}
	\hfill 
	\begin{subfigure}{0.49\linewidth}
		\centering
		\includegraphics[width=\linewidth]{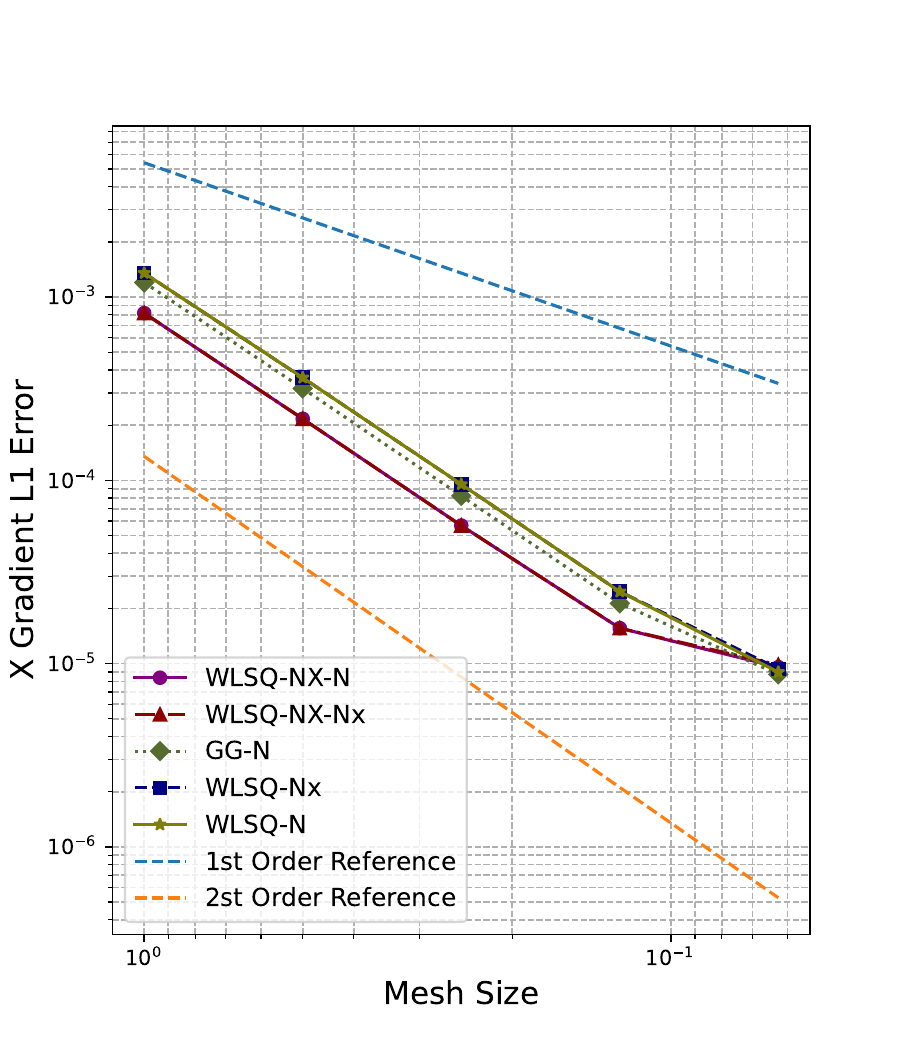}
		\caption{Accuracy of gradient reconstruction in the x-direction for a triangle cavity quadrilateral.}
	\end{subfigure}
	\begin{subfigure}{0.49\linewidth}
		\centering
		\includegraphics[width=\linewidth]{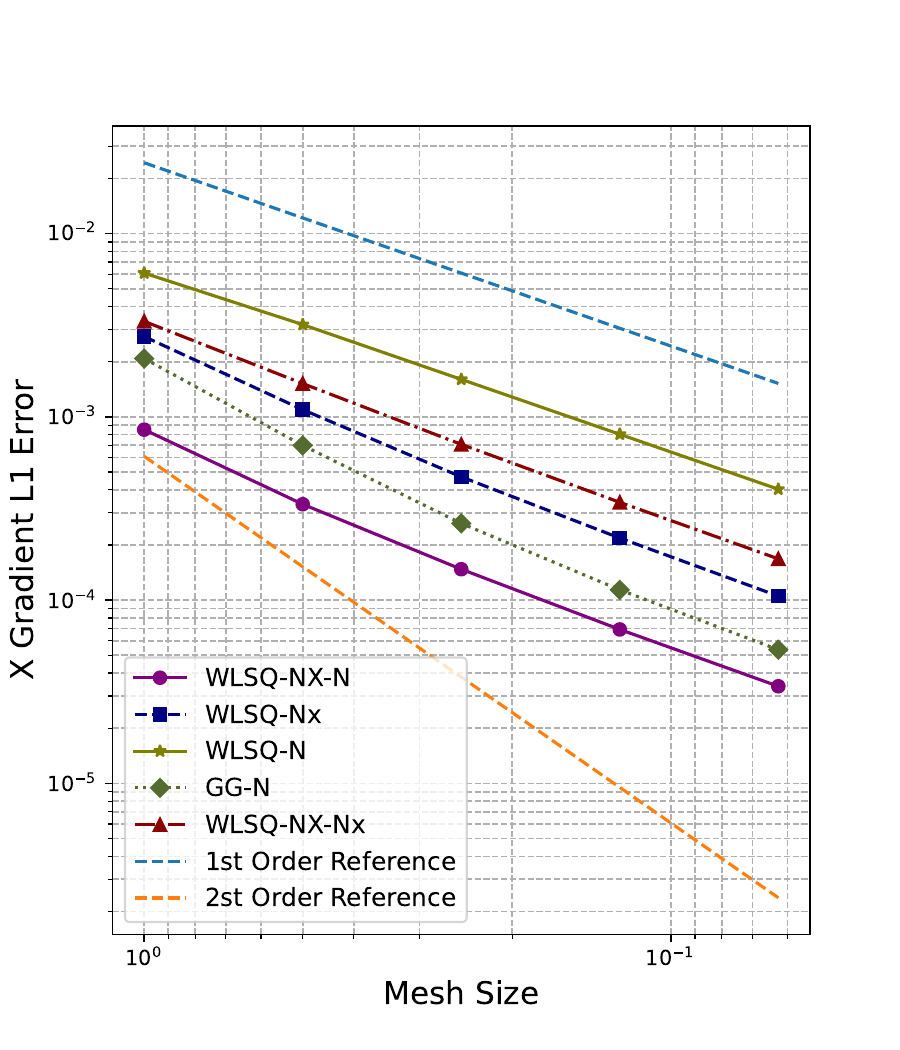}
		\caption{Accuracy of gradient reconstruction in the x-direction for a square cavity polygon.}
	\end{subfigure}
    \hfill 
	\begin{subfigure}{0.49\linewidth}
		\centering
		\includegraphics[width=\linewidth]{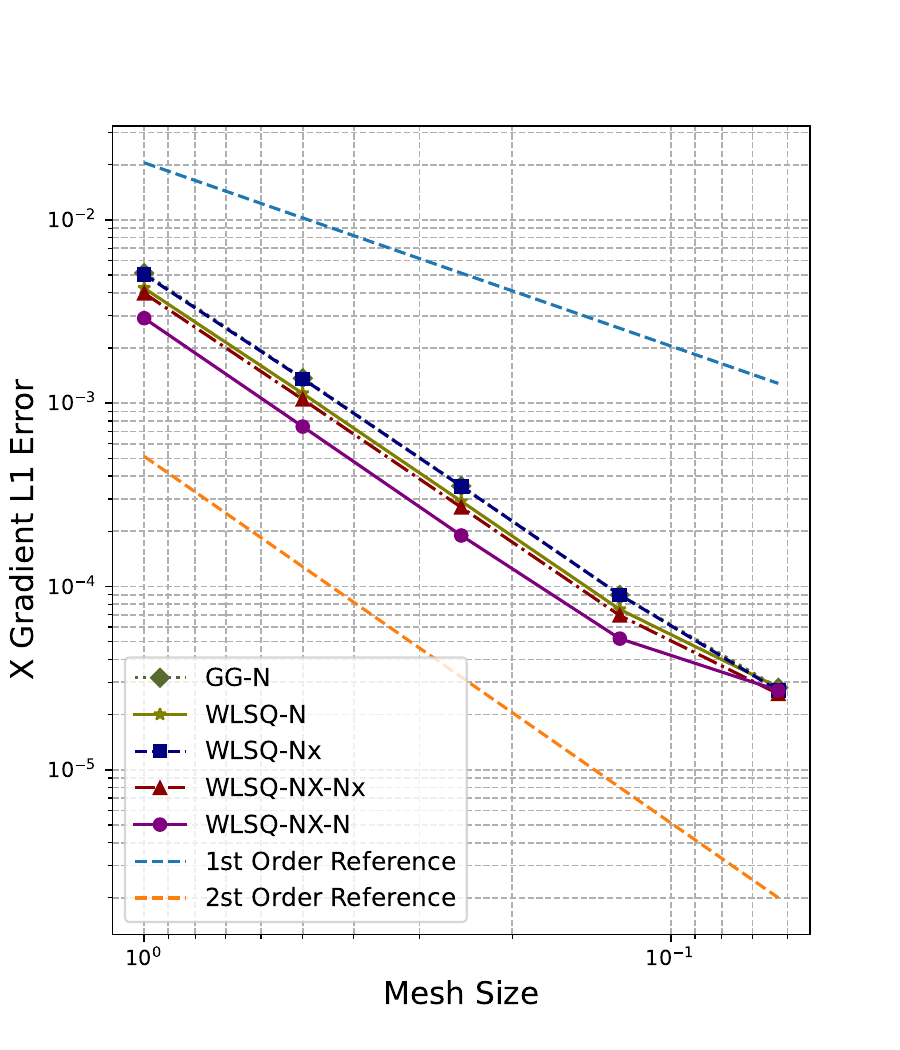}
		\caption{Accuracy of gradient reconstruction in the x-direction for an annular quadrilateral.}
	\end{subfigure}
	\caption{Gradient reconstruction error curve at grid vertices}
	\label{fig:grad_rec_node_line}
\end{figure}

By analyzing the changes in the $L_1$ norm of the $x$-direction gradient reconstruction error with respect to the mesh size, the numerical order of accuracy of the gradient reconstruction method and the size of the discretization error can be determined. It is important to note that errors can also be statistically analyzed using the $L_2$ or $L_{\infty}$ norms, which would yield similar results. However, this paper only presents the $L_1$ norm analysis results. Fig.\ref{fig:grad_rec_node_line} shows the grid convergence curves of the $x$-direction gradient errors for different gradient reconstruction methods at the grid points. The terms "1st order ref." and "2nd order ref." in the figure represent first-order and second-order precision reference curves, respectively, providing reference slopes for first and second-order precision curves.

From Fig.\ref{fig:grad_rec_node_line}, it can be seen that when calculating the gradient reconstruction error at vertices, all the gradient reconstruction algorithms used in this subsection generally achieve second-order accuracy. However, on polygonal meshes, the accuracy of all methods reduces to first order. We believe this reduction in accuracy is due to the small template size resulting from having only three neighboring grid points on polygonal meshes. It is particularly noteworthy that, looking at the results from four meshes, the error in the quadratic gradient reconstruction is significantly lower than that of just one WSLQ gradient reconstruction. The various algorithms validated in this subsection also generally possess second-order accuracy on triangular meshes, which is due to the advantageous arrangement of our variables. This is because, on triangular meshes, the number of first-order neighboring vertices at a vertex is obviously greater than the number of first-order neighboring cells at a cell.

\begin{figure}[!htbp]
	\centering
	\begin{subfigure}{0.49\linewidth}
		\centering
		\includegraphics[width=\linewidth]{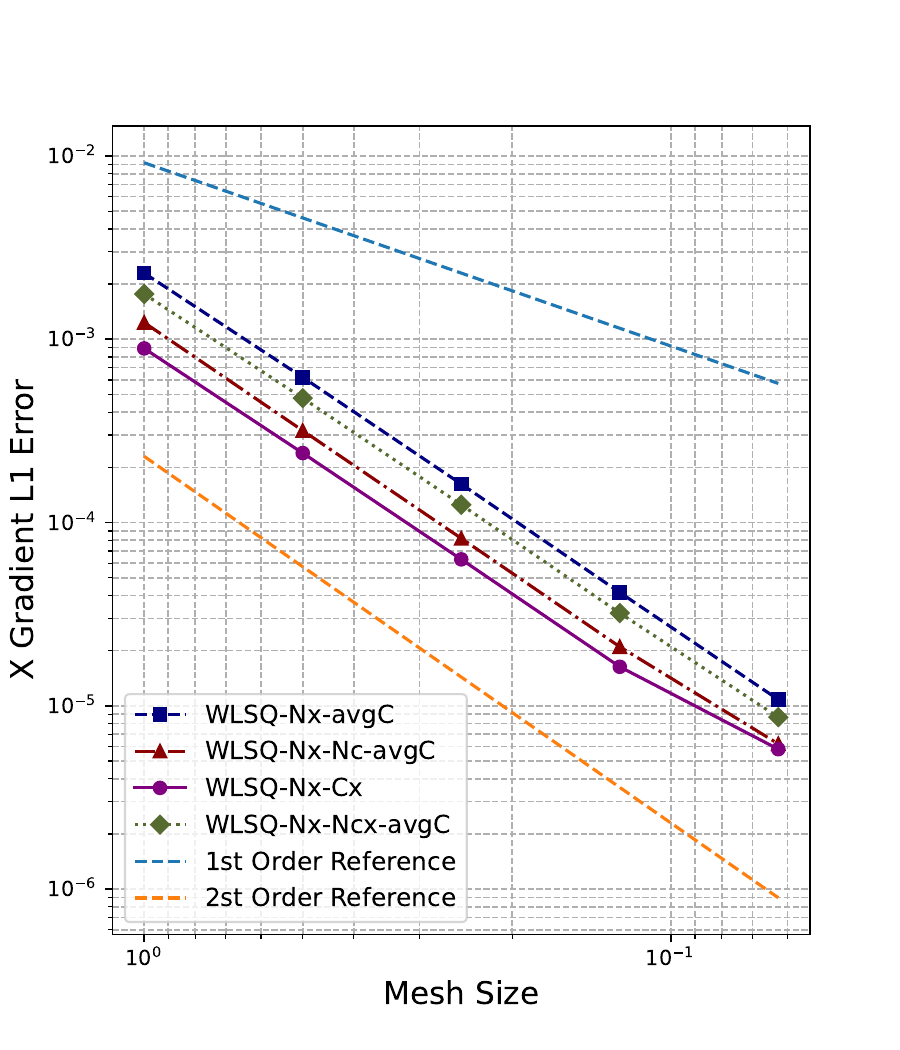}
		\caption{Accuracy of gradient reconstruction in the x-direction for a square cavity quadrilateral.}
	\end{subfigure}
	\hfill 
	\begin{subfigure}{0.49\linewidth}
		\centering
		\includegraphics[width=\linewidth]{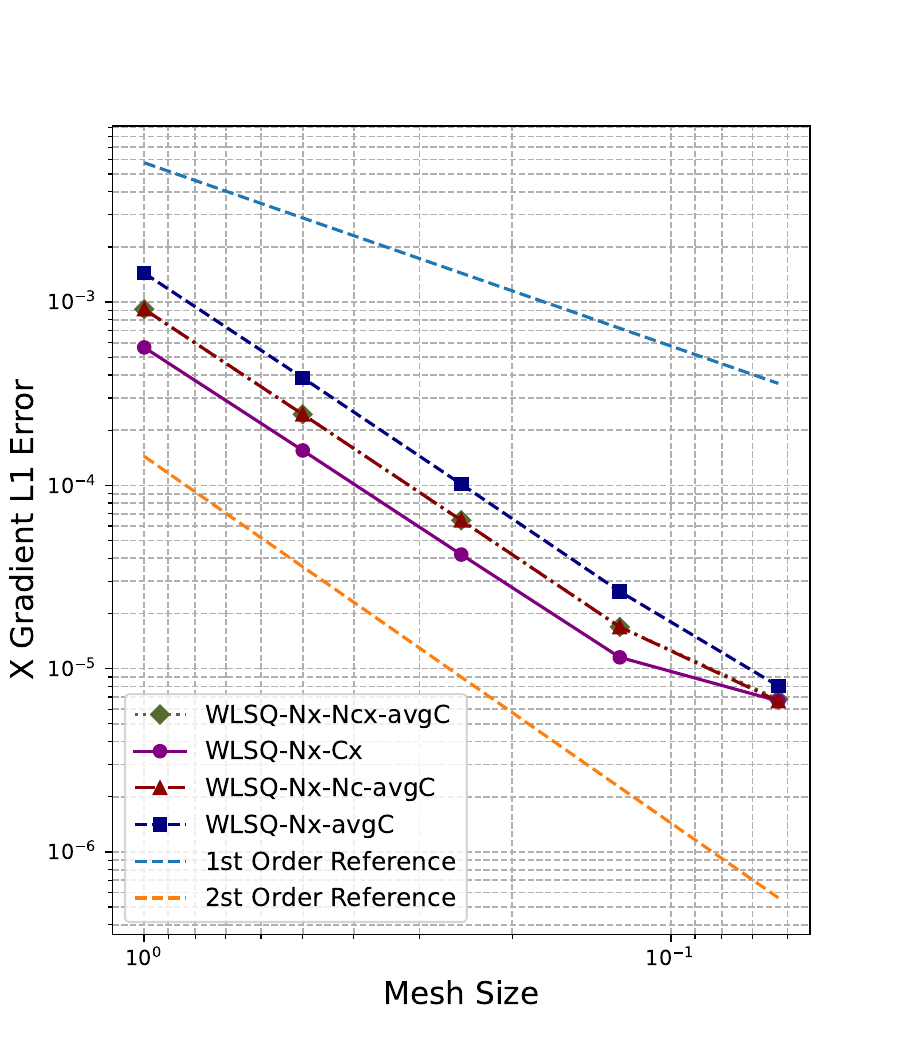}
		\caption{Accuracy of gradient reconstruction in the x-direction for a triangle cavity quadrilateral.}
	\end{subfigure}
	\begin{subfigure}{0.49\linewidth}
		\centering
		\includegraphics[width=\linewidth]{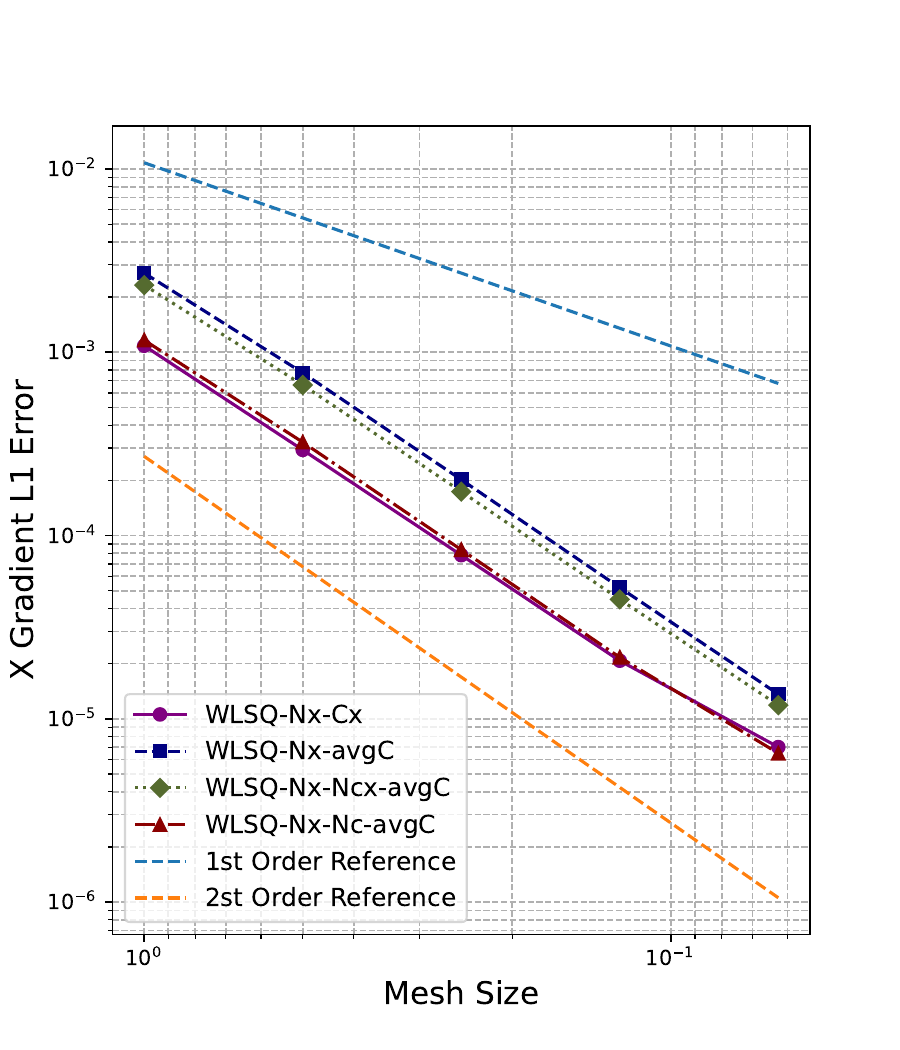}
		\caption{Accuracy of gradient reconstruction in the x-direction for a square cavity polygon.}
	\end{subfigure}
    \hfill 
	\begin{subfigure}{0.49\linewidth}
		\centering
		\includegraphics[width=\linewidth]{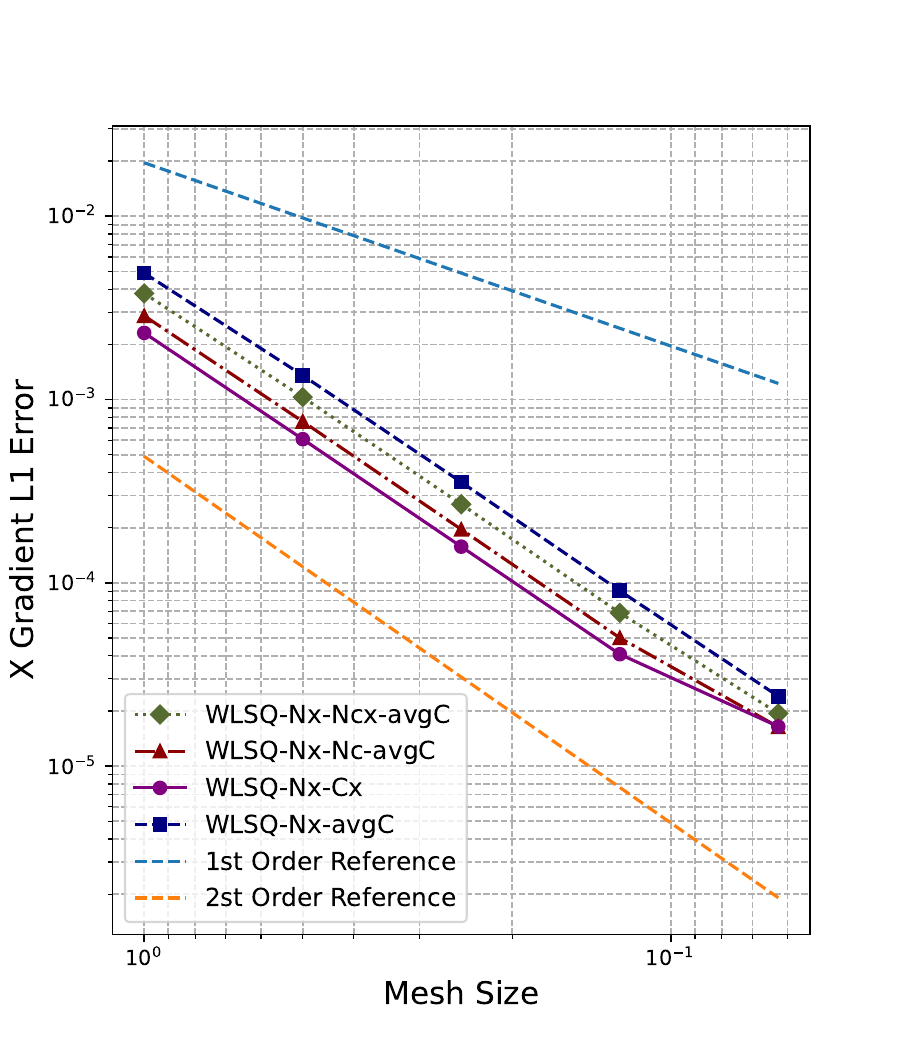}
		\caption{Accuracy of gradient reconstruction in the x-direction for an annular quadrilateral.}
	\end{subfigure}
 
	\caption{Gradient reconstruction error curve at the centroids of the grid cells.}
	\label{fig:grad_rec_cell_line}
\end{figure}

\section{GPU-Adapted Data Structures in Gen-FVGN}\label{sec:Gen-FVGN-datastructure}

In our investigations, Gen-FVGN is adept at adapting to arbitrary unstructured grids, inclusive of those comprising both triangles and quadrilaterals. In conventional CPU-based FVM methods, values on edges/faces or cells are typically iterated over each face using a for loop, leading to the reconstruction of physical quantities on the face. Evidently, the usage of for loops becomes non-viable on highly parallel GPUs. We necessitate a vectorized approach, rooted in SIMD (Single Instruction Multiple Data) paradigms, ensuring that all our computational operations remain differentiable, thus allowing derivatives with respect to learnable parameters \cite{list2022learned}. Given these considerations, we introduce five index variables to facilitate vectorized computations on unstructured grids.

Following the previous discussions, we have theoretically introduced the various processing steps of Gen-FVGN. However, the support for this algorithm must be provided by differentiable data structures. Therefore, this section primarily discusses how the variables used in the discrete processes related to the finite volume method in Gen-FVGN are implemented using graph data structures for storage and computation.

\begin{figure}[!ht]
\begin{minipage}{\linewidth}
    \centering
    \includegraphics[width=1\textwidth]{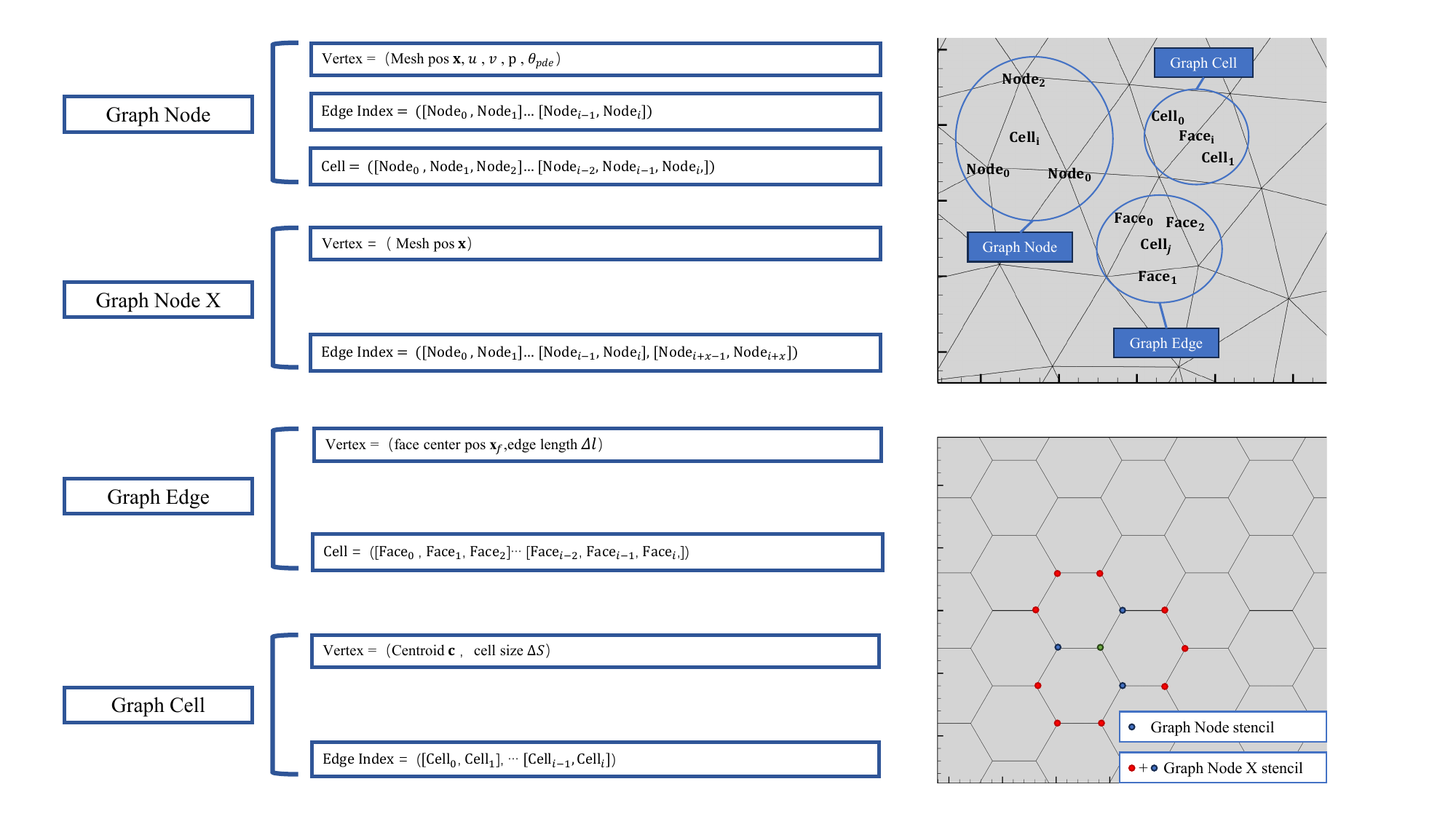}
    \caption{Grad template storage structure based on graph data structure}
    \label{fig:graph_data_structure}
\end{minipage}
\end{figure}

From Fig.\ref{fig:graph_data_structure}, it can be seen that the main structure of Gen-FVGN consists of four graphs. These are named Graph Node ($G_n$), Graph Node X ($G_{nx}$), Graph Edge ($G_e$), and Graph Cell ($G_c$). Each graph contains two to three attributes: vertex attributes, edge sets, and cell sets. The vertex attributes contain various data stored for the vertices of the graph, such as the mesh vertex coordinates stored in the vertex attributes of $G_n$ and $G_{nx}$. The edge set stores the connectivity relationships between the vertices of the graph. As shown on the right side of Fig.\ref{fig:graph_data_structure}, for $G_c$, its edge set stores the indices of the cells, thus representing the adjacency relationships between cells. The cell set attribute indicates the cells formed by the vertices in the graph. It can be seen that only the graphs $G_n$, $G_e$, and $G_c$ possess a cell set. They respectively represent the mesh cells composed of certain vertices (Cells Node), the current cell composed of certain edges or faces (Cells Face), and the cell indices corresponding to the first two cell sets (Cells Index).

\subsection{Integration Process}

\begin{figure}[!ht]
\begin{minipage}{\linewidth}
    \centering
    \includegraphics[width=1\textwidth]{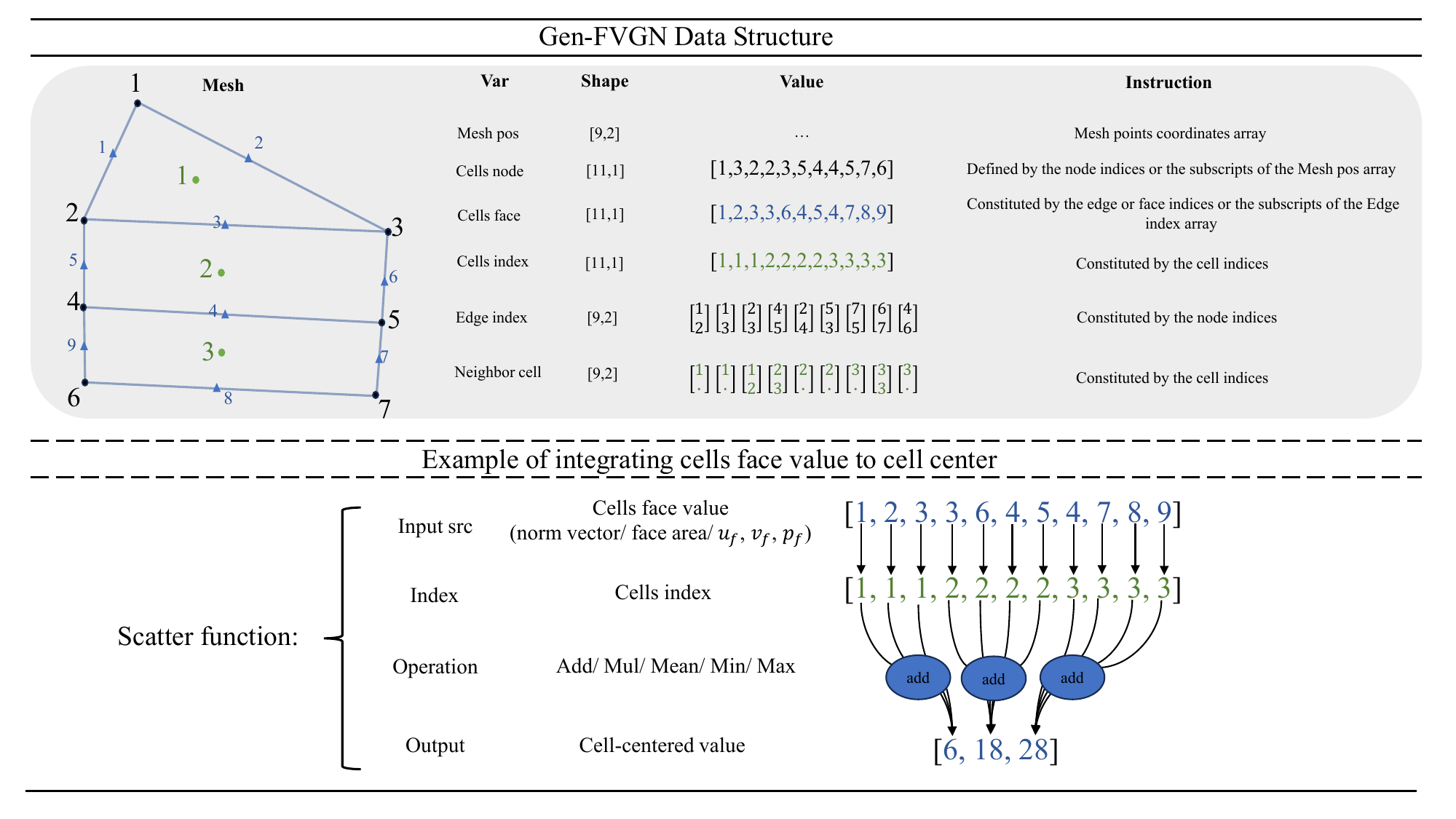}
    \caption{Data structure of Gen-FVGN and the integral calculation algorithm capable of parallel execution on GPUs}
    \label{fig:Gen-FVGN_data_struture}
\end{minipage}
\end{figure}

After introducing the Gen-FVGN data structure, we will now discuss in detail how to utilize these data structures to implement the integration calculations of the finite volume method in parallel on a GPU. Specifically, we demonstrate this process through the implementation of two key algorithms:

\quad 1. Accumulation of surface physical quantities to the cell center: As shown in Fig.\ref{fig:Gen-FVGN_data_struture}, in the process of performing integration calculations, i.e., the application of the divergence theorem, Gen-FVGN first needs to handle the cell set stored in $G_c$ and the surface physical quantities located at the vertices stored in $G_e$. Through a Pytorch-based Scatter function\cite{fey2019fast}, the algorithm efficiently accumulates the physical quantity values of each cell surface to the corresponding cell center. The key to this step lies in the parallel execution capability of the Scatter function, which allows simultaneous processing of multiple cells and surface quantities, greatly enhancing the computational efficiency.

\quad 2. Optimization and time complexity of parallel computation: In an ideal situation, this accumulation process can achieve a time complexity of O(1), meaning that the execution time remains almost constant regardless of the scale of the computation. This is thanks to the powerful parallel processing capabilities of the GPU and the optimization of the algorithm design, ensuring that even when dealing with large-scale fluid dynamics simulations, high computational performance is maintained.

\begin{figure}[!htbp]
\begin{minipage}{\linewidth}
    \centering
    \includegraphics[width=1\textwidth]{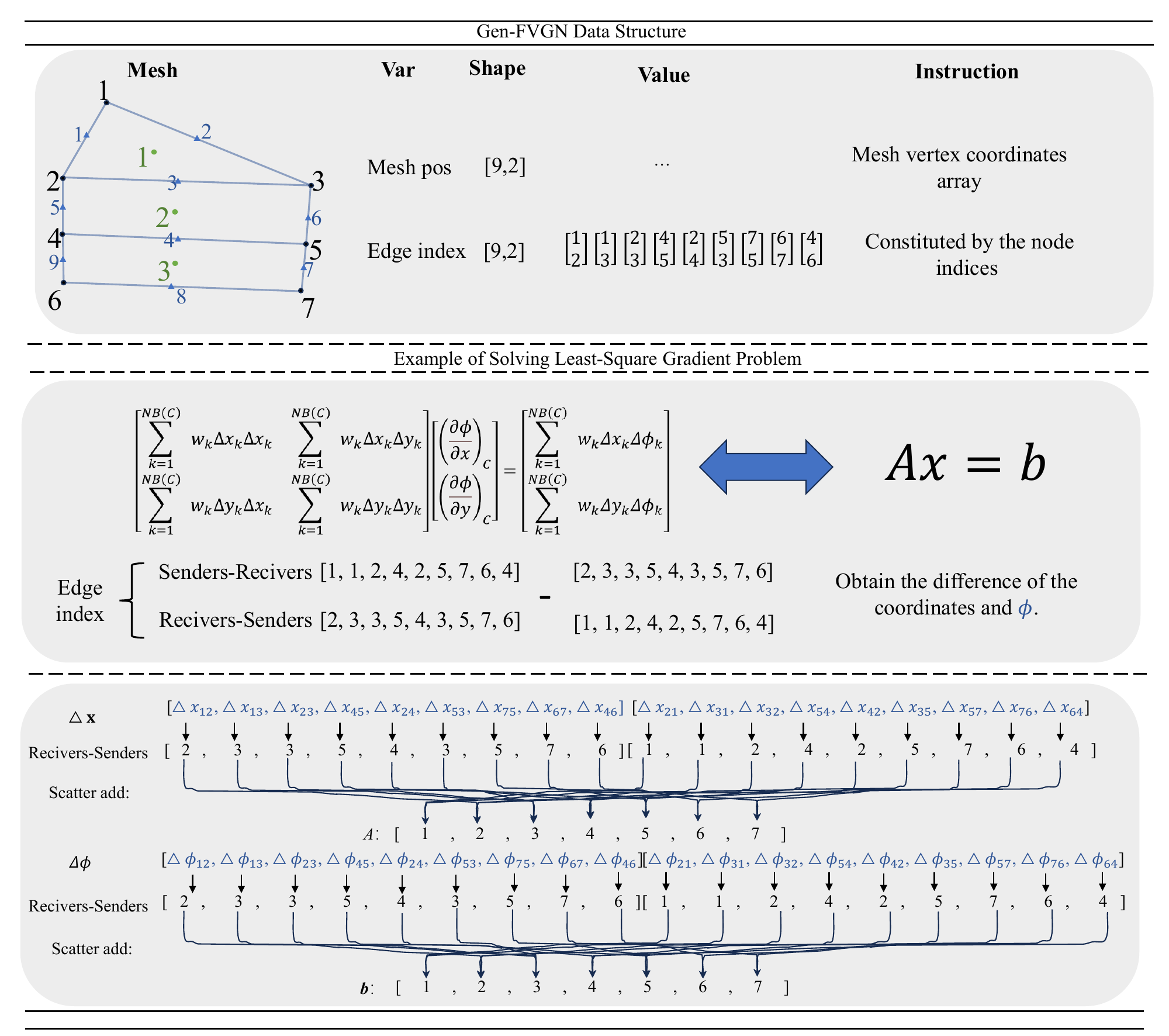}
    \caption{A Vertex-based Weighted Least Squares (WLSQ) gradient reconstruction algorithm that can be parallelized on GPUs.}
    \label{fig:WLSQ_example}
\end{minipage}
\end{figure}

Through the implementation of these two algorithms, we demonstrate how to effectively use the Gen-FVGN data structure and GPU parallel computing capabilities to perform the integration calculation steps in the finite volume method. This not only speeds up the computation but also provides reliable technical support for the efficient implementation of complex fluid dynamics simulations.

\subsection{Implementation Process of the Least Squares Method}

In the implementation process of the Weighted Least Squares (WLSQ), we first focus on the accuracy and computational efficiency of the gradient reconstruction. As shown in Fig.\ref{fig:WLSQ_example}, our algorithm, Gen-FVGN, takes the following steps for gradient reconstruction:

1. Edge set and vertex coordinate interpolation: The algorithm first uses the edge set stored in $G_n$ to construct coordinate differences $\bigtriangleup x$ and corresponding physical quantity differences $\bigtriangleup \phi$ for vertices at both ends of each edge. This step is the foundation of gradient reconstruction, ensuring the accurate representation of the spatial variation of physical quantities.

2. Parallel accumulation operation: Subsequently, through the Scatter function and indexed by Receivers-Senders pairs, the algorithm parallelly accumulates the $\bigtriangleup x$ and $\bigtriangleup \phi$ of each edge. This parallel computing step significantly improves processing speed and is key to utilizing the parallel processing capabilities of GPUs.

3. Building the least squares equation set: Through the above steps, we obtain the coefficient matrix $\mathbf{A}$ and the constant vector $\mathbf{b}$ for the equation set used in least squares. These elements are necessary conditions for solving the least squares problem and obtaining vertex gradients.

4. Gradient solution: Finally, using the matrix solving interface provided by the deep learning framework, the algorithm calculates the gradient at each vertex. This step not only utilizes the efficiency of modern deep learning frameworks but also ensures the accuracy of gradient calculations.

Through these steps, our Weighted Least Squares method can not only be efficiently parallelized on GPUs but also accurately reconstructs gradients at each vertex, providing a powerful mathematical tool for fluid dynamics simulations.

\section{All Trained Mesh Preview}
In this section, the meshes we used include triangular meshes, quadrilateral meshes, and hybrid meshes of triangles and quadrilaterals. Some of the meshes feature boundary layer refinement settings to verify the solution capabilities of Gen-FVGN when dealing with anisotropic meshes, as mentioned in Sec.\ref{sec:numerical_validate}.
\begin{figure}[!htbp]
\centering
\begin{minipage}{\linewidth}\centering
    \includegraphics[width=1\textwidth]{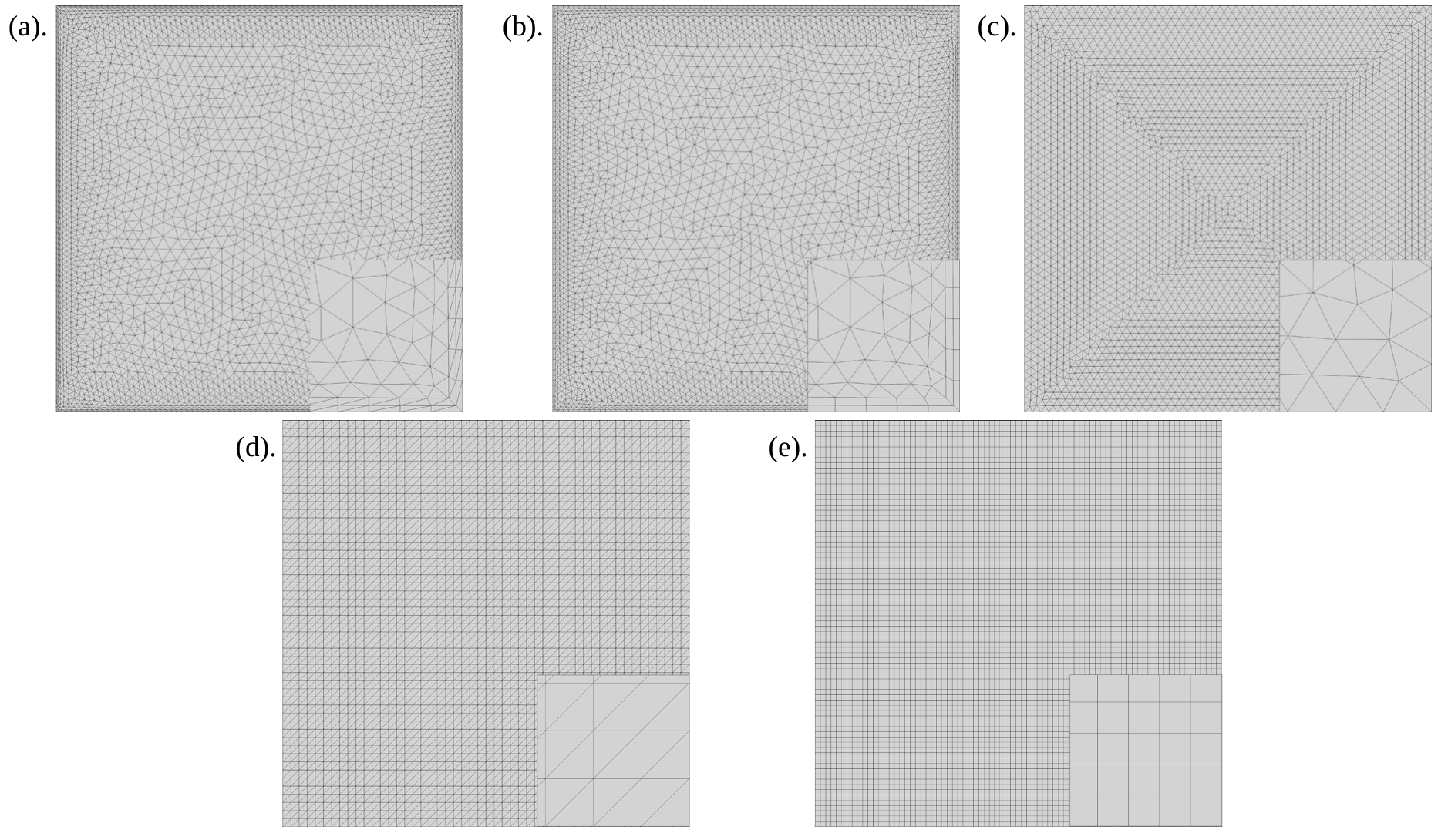}
    \end{minipage}
    \caption{Cavity grids: (a) Anisotropic triangular mesh; (b) Triangular-quadrilateral hybrid mesh; (c) Triangular mesh generated by the advancing front method; (d) Isotropic triangular mesh; e) Isotropic quadrilateral mesh}
    \label{fig:cavity_mesh}
\end{figure}

It should be noted that the overall domain width of all the square cavity meshes shown in Fig.\ref{fig:cavity_mesh} is $1[m]$. The first two meshes include wall-refined meshes. We will perform validation of the solutions for the Poisson equation and the wave equation on these five meshes.

\begin{figure}[!htbp]
\centering
\begin{minipage}{\linewidth}\centering
    \includegraphics[width=0.8\textwidth]{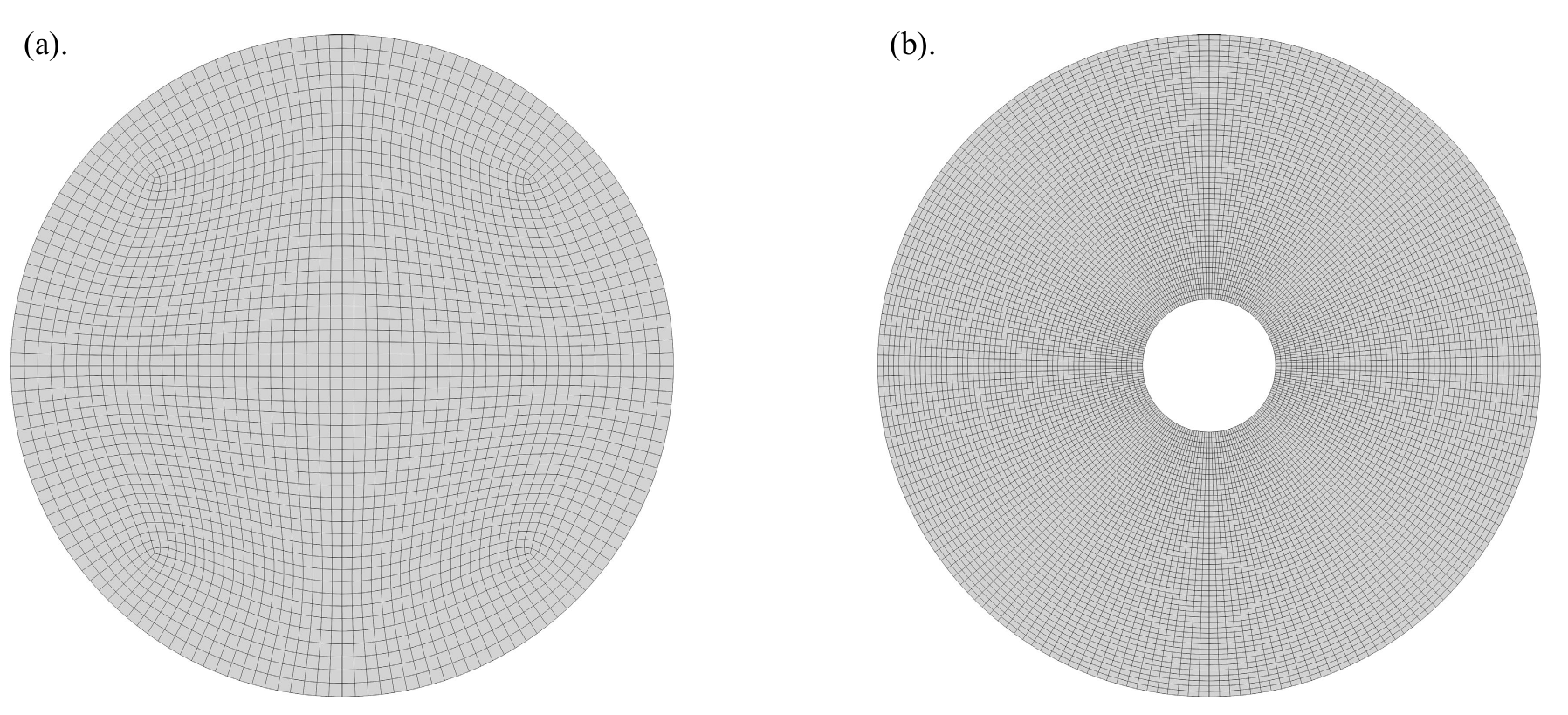}
    \end{minipage}
    \caption{Circular and annular meshes; (a) Free quadrilateral mesh; (b) Isotropic quadrilateral mesh.}
    \label{fig:circular_ring_mesh}
\end{figure}

\begin{figure}[!htbp]
\centering
\begin{minipage}{\linewidth}\centering
    \includegraphics[width=1\textwidth]{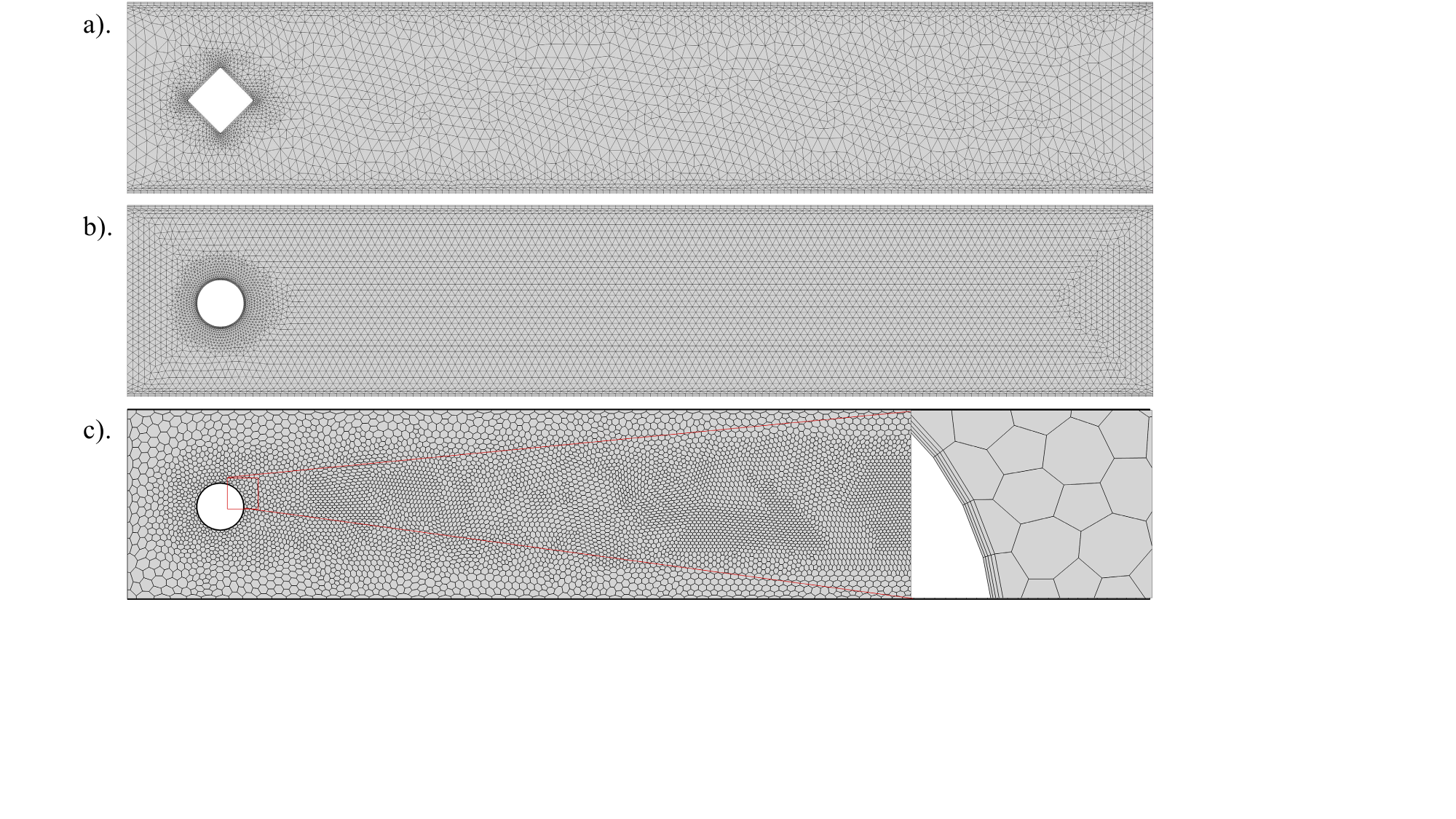}
    \end{minipage}
    \caption{Pipe flow case meshes; (a) Square column triangle-quadrilateral hybrid mesh; (b) Cylinder triangle-quadrilateral hybrid mesh; (c) Cylinder polygon-quadrilateral hybrid mesh.}
    \label{fig:pipeflow_mesh}
\end{figure}

\begin{figure}[!htbp]
\centering
\begin{minipage}{\linewidth}\centering
    \includegraphics[width=1\textwidth]{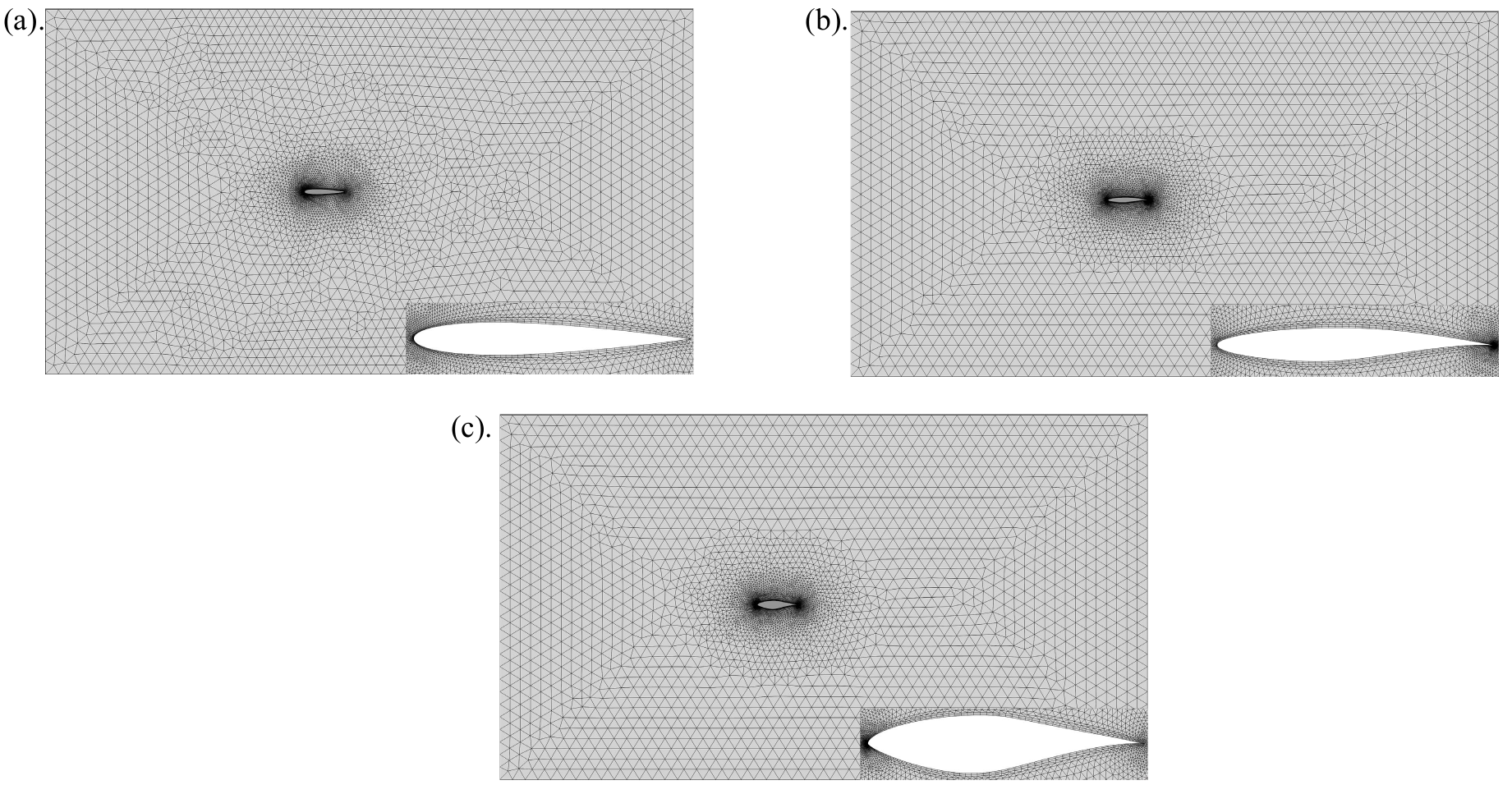}
    \end{minipage}
    \caption{Outflow field meshes; (a) NACA0012 airfoil triangle-quadrilateral hybrid mesh; (b) RAE2822 airfoil triangle-quadrilateral hybrid mesh; (c) S809 airfoil triangle-quadrilateral hybrid mesh; (d) Cylinder free quadrilateral mesh.}
    \label{fig:farfield_mesh}
\end{figure}

The number of elements in the aforementioned meshes was not specifically set. For the specific problems addressed, the number of elements provided here may not be able to accurately capture the flow field information corresponding to the Reynolds numbers used for training. In other words, we did not perform grid independence verification during the training and solution process. However, this paper focuses on the effectiveness of the method, and thus our main consideration is to verify the generalization ability of the network. Clearly, narrowing the range of Reynolds numbers used for training can avoid this issue. This is because we can perform a grid independence verification using a commercial solver first, and then begin our model training.

\end{document}